\documentclass[%
 reprint,
 amsmath,amssymb,
 aps,
]{revtex4-1}

\usepackage{multirow}
\usepackage{bm}
\usepackage{graphicx}
\usepackage{amsmath}
\usepackage[colorlinks=true, allcolors=blue]{hyperref}
\usepackage{jabbrv}
\newcommand\Tstrut{\rule{0pt}{3.5ex}}        
\newcommand\Bstrut{\rule[-2.1ex]{0pt}{0pt}}

\begin{document}

\preprint{APS/123-QED}

\title{Nuclear shell-model simulation in digital quantum computers}

\author{A. P\'{e}rez-Obiol}
\email{axel.perezobiol@bsc.es}
\affiliation{Barcelona Supercomputing Center, 08034 Barcelona, Spain}

\author{A. M. Romero}
\email{a.marquez.romero@fqa.ub.edu}
\author{J. Men\'{e}ndez}
\email{menendez@fqa.ub.edu}
\author{A. Rios}
\email{arnau.rios@fqa.ub.edu}
\affiliation{Departament de Física Quàntica i Astrofísica (FQA), Universitat de Barcelona (UB), c. Martí i Franqués, 1, 08028, Barcelona, Spain\\
Institut de Ciències del Cosmos (ICCUB), Universitat de Barcelona (UB), c. Martí i Franqués, 1, 08028 Barcelona, Spain}

\author{A. Garc\'{i}a-S\'{a}ez}
\email{artur.garcia@bsc.es}
\affiliation{Barcelona Supercomputing Center, 08034 Barcelona, Spain\\
Qilimanjaro Quantum Tech, 08007 Barcelona, Spain}

\author{B. Juli\'{a}-D\'{i}az}
\email{bruno@fqa.ub.edu}
\affiliation{Departament de Física Quàntica i Astrofísica (FQA), Universitat de Barcelona (UB), c. Martí i Franqués, 1, 08028 Barcelona, Spain\\
Institut de Ciències del Cosmos (ICCUB), Universitat de Barcelona (UB), c. Martí i Franqués, 1, 08028 Barcelona, Spain}

\date{\today}

\begin{abstract}
The nuclear shell model is one of the prime many-body methods to study the structure of atomic nuclei, 
but it is hampered by an exponential scaling on the basis size as the number of particles increases.
We present a shell-model quantum circuit design strategy to find nuclear ground states by exploiting an adaptive variational quantum 
eigensolver algorithm.
Our circuit implementation is in excellent agreement with classical shell-model simulations for
a dozen of light and medium-mass nuclei, including neon and calcium isotopes. We quantify the circuit depth, width and number of gates to encode realistic 
shell-model wavefunctions. Our strategy also addresses explicitly energy measurements and the required number of circuits to perform them. Our simulated circuits approach the benchmark results exponentially with a polynomial scaling in quantum resources for each nucleus. 
This work paves the way for quantum computing shell-model studies across the nuclear chart and our quantum resource quantification may be used in configuration-interaction calculations of other fermionic systems.
\end{abstract}

\maketitle

Atomic nuclei are complex many-body systems formed by protons and neutrons (collectively denoted as nucleons) bound by the strong nuclear force. Nuclei exhibit captivating properties such as the coexistence of spherical and deformed shapes at low energies~\cite{Taniuchi:2019pen,Butler:2019qox,Tsunoda:2020gpt}, strong short-range correlations between pairs of nucleons~\cite{CLAS:2020mom}, or decay modes driven by the strong~\cite{Mukha06}, weak~\cite{Hinke12} or electromagnetic~\cite{Walz15} forces. Furthermore, nuclear decays are crucial to understand the origin of heavy elements in the universe~\cite{Cowan:2019pkx}, and experiments using nuclei aim to answer fundamental physics questions such as which is the nature of dark matter~\cite{Aalbers:2022dzr}, why matter dominates over antimatter in the universe~\cite{Engel:2013lsa}, or whether neutrinos are their own antiparticles~\cite{Avignone:2007fu}.

The nuclear shell model, also known as the configuration interaction method, is one of the leading many-body approaches to study the structure of nuclei. The shell model is grounded in the idea that, in a similar fashion to electrons in an atom, nucleons occupy orbitals organized in \emph{shells} of different energies~\cite{mayerII,Haxel:1949fjd}. Nuclear states are then obtained by computationally intensive diagonalizations of the nuclear Hamiltonian in a many-body configuration space comprising one or several shells. In spite of impressive progress in recent decades~\cite{brown1988status, Caurier:2004gf,otsuka2020evolution,Stroberg:2019mxo}, the exponential scaling of the many-body Hilbert space with the number of nucleons ultimately prevents the application of the shell model across the entire nuclear chart, particularly in heavy nuclei.

Quantum computing promises to circumvent limitations
associated to any exponentially-scaling many-body system using the principle of superposition of qubit states~\cite{arute2019quantum}.
In the current noisy intermediate-scale quantum (NISQ) device era~\cite{preskill2018quantum}, \emph{variational quantum eigensolvers} (VQE)~\cite{peruzzo2014variational, mcclean2016theory} are among the most successful algorithms~\cite{bharti2022noisy} exploiting 
the benefits of quantum computing to deal with complex many-body problems in physics~\cite{Cerezo2021,Tilly2022} 
and chemistry~\cite{uccrev,McArdle2020,haidar2022open}.

Quantum many-body systems that have been used as VQE testbeds include the Fermi-Hubbard~\cite{fh1}, Ising~\cite{cervera2018exact} and Lipkin-Meshkov-Glick models~\cite{cervia2021lipkin,harsha2018difference,faba3,wahlen2017merging,Robin2023}, superfluid systems~\cite{ph1,ph2}, hadrons~\cite{Qian:2021jxp} or molecules~\cite{grimsley2019adaptive,sapova22,Feniou2023}.

\begin{figure*}[t]
     \centering
     \includegraphics[width=\linewidth]{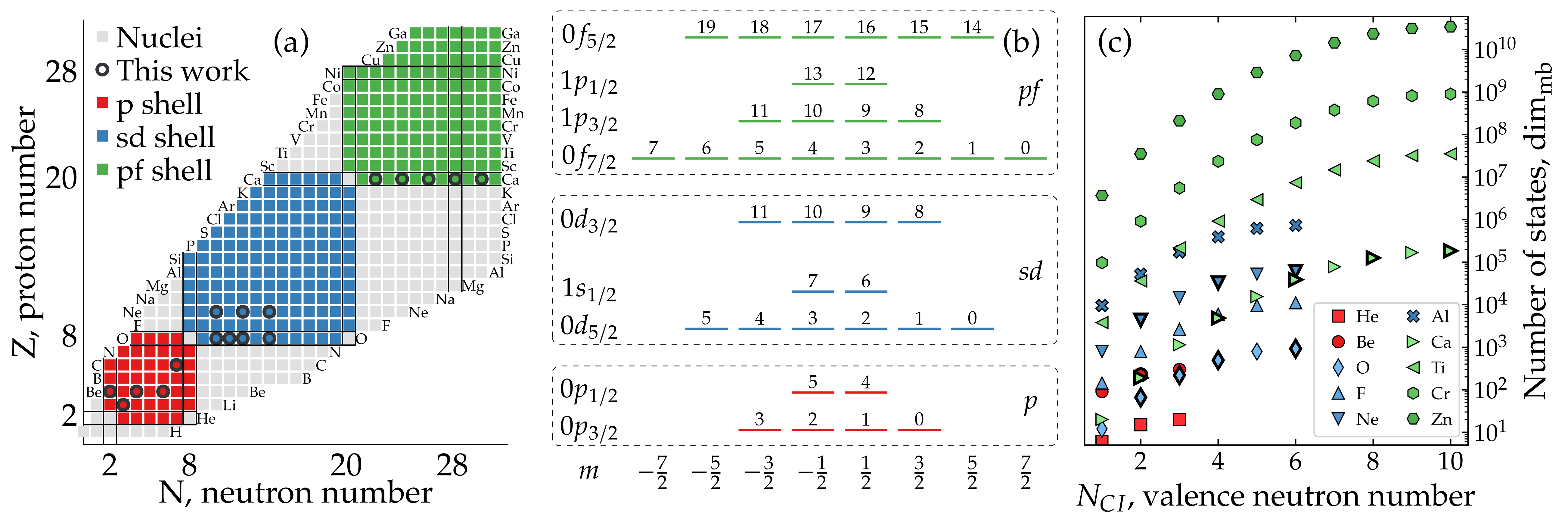}
     \caption{     
     {\bf The shell model and quantum encoding.} 
     Panel (a): Segr\`e chart covering the $p$, $sd$ and part of the $pf$ shell. Solid lines indicate neutron and proton magic numbers. Open circles show the isotopes studied in this work.
     Panel (b): schematic representation of the $p$-, $sd$- and $pf$-shell configuration spaces. 
     The number on top of every single-particle state is the qubit label for the implementation in a quantum device under a Jordan-Wigner mapping. 
     Panel (c): number of many-body configurations, $\dim_{\rm{mb}}$, in the $M$-basis as a function of the number of active neutrons in the configuration space, $N_{\rm{CI}}$. 
     We show results for the isotopic chains of He and Be in the $p$ shell; O, F, Ne, and Al in the $sd$ shell; and Ca, Ti, Cr, and Zn in the $pf$ shell. 
     Isotopes beyond the middle of the shell are not shown since the number of configurations is symmetric. Bold marker lines highlight nuclei studied in this work.}
     \label{fig:master}     
\end{figure*}

In general, a VQE implementation requires a series of well-defined stages~\cite{Tilly2022}, involving
a) a mapping between physical degrees of freedom (eg fermionic operators) and the qubits in a quantum computer;
b) the preparation of an initial reference state; 
c) a (potentially iterative) variational optimization; 
d) a measurement strategy for expectation values of operators (most importantly, the Hamiltonian);
and e) an error mitigation scheme. 
Previous nuclear shell-model studies have only partially tackled these problems~\cite{Dumitrescu2018,LuKlco2019,stetcu,papenbrock}.
The aim of this article is to present a circuit design strategy that explicitly addresses all these aspects
to solve the nuclear shell model in a quantum computer. 
We 
also quantify the necessary 
circuit resources, such as depths and widths, to achieve precise predictions for
nuclear masses. 
We do this in a set of test nuclei across different nuclear shells.
To this end, we perform (classical) baseline 
simulations on the corresponding circuit architectures  and benchmark the results 
against diagonalizable 
shell-model simulations as well as independent ADAPT-VQE simulations without 
an explicit circuit implementation.

\section*{Results}

\subsection*{Nuclear shell model}\label{sec:nsm}

The nuclear shell model~\cite{brown1988status, Caurier:2004gf,otsuka2020evolution,Stroberg:2019mxo} considers nuclei composed by an inert core of nucleons, which do not explicitly contribute to the dynamics, and a set of valence protons and neutrons interacting in a relatively small configuration space. This space is usually bounded by two \emph{magic numbers}, which denote special configurations of protons or neutrons leading to particularly stable nuclei. Magic numbers thus define shells with large energy gaps between them. Configuration spaces used in shell-model calculations usually comprise one or two shells.
Panel (a) of Fig.~\ref{fig:master} shows the light to mid-mass region of the isotope chart.
We highlight areas where the $p$, $sd$ and $pf$ shell-model calculations are routinely employed. 

Since the nuclear force is rotationally invariant and nucleons are fermions, it is useful to work in a single-particle basis with states with quantum numbers $n\,l_{j}$, where $n$ is the principal quantum number, $l$ the orbital angular momentum and $j$ the total angular momentum~\cite{Shalit1963}.
This basis also includes $m$ third-component projections of $j$ degenerate in energy. 
The nuclear Hamiltonian is also to a very good approximation the same for neutrons and protons, so it is customary to define, additionally, the isospin quantum number $t=1/2$, with third component $t_z$ discerning protons and neutrons~\cite{Talmi1993}. Many-body nuclear states have good total angular momentum $J$ and isospin $T$, with respective third components $M$ and $T_z$ given by the sum of the third components of all nucleons in the nucleus~\cite{varshalovich1988quantum}.

The nuclear Hamiltonian in a given configuration space
can be written as
\begin{equation}\label{eq:smham}
    H_{\rm{eff}} = \sum_i \varepsilon_i a_i^{\dag} a_i + \frac{1}{4} \sum_{ijkl}\bar{v}_{ijkl}
    a_i^{\dag} a_j^{\dag} a_l a_k \,,
\end{equation}
where $\varepsilon_i$ is the energy of the single-particle state $i$ and $\bar{v}_{ijkl} = v_{ijkl} - v_{ijlk}$ are antisymmetrized two-body matrix elements. 
$a_i$ and $a_i^{\dag}$ are fermionic annihilation and creation operators 
associated to each  single-particle state, $i$. 
The matrix elements $\bar{v}_{ijkl}$ can be obtained~\cite{HjorthJensen1995,Stroberg:2019mxo} 
from an effective 
field theory of the underlying theory of the nuclear force, quantum chromodynamics~\cite{Epelbaum:2008ga}. Here, instead, we use 
standard phenomenological Hamiltonians, with components adjusted to better reproduce key 
properties of selected nuclei~\cite{Poves:1981zz}. We choose the 
Cohen-Kurath interaction in the $p$ shell~\cite{cohen1965effective}, 
USDB in the $sd$ shell~\cite{Brown2006} and 
KB3G in the $pf$ shell~\cite{Poves2001}.

A suitable many-body basis, also referred to as Fock space, for shell-model calculations is provided by 
the so-called $M-$scheme~\cite{Talmi1993},
in which the Slater determinant states are chosen to have a 
well-defined $M$. 
$T_z=(N-Z)/2$ is also well defined because the number of neutrons $N$ and protons $Z$ is fixed. Nuclear states are thus expanded in this basis,
\begin{equation}
   | JM\,TT_z\rangle = \sum_{\alpha} c_{\alpha}  | \alpha, M T_z\rangle,
\end{equation} 
and nuclear wavefunctions and their corresponding energies are eigenvectors and eigenvalues of the Hamiltonian matrix in the basis of Slater determinants. 
The $c_\alpha$ coefficients are obtained through  
diagonalization employing state-of-the-art nuclear shell-model codes~\cite{caurier1999antoine,Shimizu:2019xcd,brown2014shell,johnson2018bigstick} 
and ensure that  eigenstates have  good $J$ and $T$ 
quantum numbers.

However, this framework faces a steep computational bottleneck in terms of the maximum size of the Hamiltonian matrix from which the lowest eigenvalues and eigenvectors can be calculated. 
The dimension of the single-particle basis of a nuclear shell consisting of several orbitals $nl_j$ is
\begin{align}
    \dim_{\rm{sp}} = \sum_j (2j+1),
    \label{eq:dimsp}
\end{align}
where the sum runs over the $j$ values in a given configuration space, see
panel (b) of Fig.~\ref{fig:master} for details. 
The corresponding number of Slater determinants grows combinatorially as 
\begin{equation}\label{eq:combpart}
    \dim_{\text{mb}} = \binom{\dim_{\text{sp}}}{N_{\text{CI}}} \times \binom{\dim_{\text{sp}}}{Z_{\text{CI}}},
\end{equation}
where $N_{\text{CI}}$ ($Z_{\text{CI}}$) is the number of active neutrons (protons) in the configuration space.
Let us consider the $sd$ shell, comprising the $1s_{1/2}$, $0d_{3/2}$ and $0d_{5/2}$ orbitals for both protons and neutrons, and the $pf$ shell, comprising the $0f_{7/2}$, $0f_{5/2}$, $1p_{3/2}$ and $1p_{1/2}$ orbitals. There are $12$ ($20$) single-particle states in the $sd$ ($pf$) shell, so that it can describe the isotopic chains of $12$ ($20$) elements with up to $12$ ($20$) valence neutrons, as shown in panel (a)  of Fig.~\ref{fig:master}. Panel (c) illustrates the exponential scaling of the number of many-body configurations, $\dim_{\rm{mb}}$, present for isotopes of elements in 
different shells. The number of basis states needed to describe two isotopes of the same element, or two elements with the same $N$ in the same shell, can differ by three or more orders of magnitude. 

In practical calculations, this number may be reduced by about an order of magnitude due to symmetry considerations, leading to a reduced number of Slater determinants, $N_{\text{SD}}$~\cite{Caurier:2004gf}. However, the scaling in either 
$\dim_{\text{mb}}$ or $N_{\text{SD}}$
ultimately places a limit in the computational resources needed to study heavy nuclei with the nuclear shell model. This refers to both the number of operations per second, or CPU time, and the memory to store all configurations. In fact, the shell-model history is closely tied to that of computation, as larger-scale calculations became feasible with the advances in computational power and refined techniques in CPUs and GPUs~\cite{Caurier:2004gf,brown1988status,otsuka2020evolution,Stroberg:2019mxo}.

\subsection*{Variational algorithm}
\label{sec:algorithm}

Here, we implement the nuclear shell model in a quantum computer following a standard Jordan-Wigner (JW) mapping~\cite{Seeley2012,romeroquantum,papenbrock,stetcu}. 
We associate each qubit with a single-particle state in the configuration space, which can 
either be empty (projection $0$) or occupied (projection $1$). 
Panel (b) of Fig.~\ref{fig:master} shows 
the mapping between single-particle states and qubits for the 
$p$ (bottom), $sd$ (central) and $pf$ shells (top panel). 
From a memory-storage perspective, a shell-model
VQE under the JW mapping only requires as many qubits as 
single-particle states in the configuration space. 
In other words, the number of qubits remains constant for all nuclei described within a given shell. 
If a VQE can be used to diagonalize the problem and is robust against errors, the approach may
provide access to much larger configuration spaces, 
currently unattainable in classical computers.

A VQE uses the Rayleigh-Ritz variational principle~\cite{ritz, rayleigh} to calculate the ground-state of a Hamiltonian starting from an initial ansatz. Our algorithm of choice is ADAPT-VQE~\cite{grimsley2019adaptive,romeroquantum,tang2021qubit,haidar2022open,Feniou2023},
 which iteratively builds a wavefunction of the form 
 \begin{equation}
    |\psi(\bm{\theta})\rangle = \prod_{k=1}^n e^{i\theta_k A_k} |{\rm ref}\rangle,
    \label{eq:adaptwf1}
\end{equation}
where $|{\rm ref}\rangle$ is an initial (reference) state of the quantum system, 
$k$ is the iteration (or layer) index, 
$A_k$ are particle-hole excitation operators,
and $\bm{\theta}=\{ \theta_i, i=1,\dots, n \}$ are a set of variational parameters. 
We stress that the adapted wavefunction in Eq.~(\ref{eq:adaptwf1}) is 
free of Trotter-Suzuki approximation errors~\cite{barkoutsos2018quantum, childs2021theory}.
This ansatz does not require decomposing an exponential map of a sum of excitation operators, as would be the case in algorithms such as UCC-VQE~\cite{uccrev,papenbrock}.

The minimization of the energy of this wavefunction with respect to the parameters $\bm{\theta}$,
\begin{equation}
    E = \min_{\bm{\theta}} \frac{\langle \psi(\bm{\theta}) | H_{\rm{eff}} |\psi(\bm{\theta})\rangle}{\langle \psi(\bm{\theta}) |\psi(\bm{\theta})\rangle}, 
\end{equation}
can be performed classically~\cite{pellow2021comparison}
and yields an approximate ground-state energy.
Here, we use the BFGS optimiser with a gradient tolerance
set to $10^{-6}$ at every iteration.
At each layer $k$ of the iterative procedure,
the ansatz grows by one parametrized unitary, $|\psi(\bm{\theta})\rangle\to e^{i\theta_k A_k}|\psi(\bm{\theta})\rangle$.
The new operator $A_k$ is selected according to the largest energy gradient computed as
\begin{equation}
    \left. \frac{\partial E^{(n)}}{\partial \theta_k } \right\vert_{\theta_k=0} = 
    \left. i \langle \psi (\bm{\theta})| [H_{\rm{eff}},A_k] |\psi (\bm{\theta})\rangle \right\vert_{\theta_k=0} . 
    \label{eq:gradient}
\end{equation}
Thus, at every layer, the wavefunction adapts to the new information acquired in the previous optimization.
The set of parameters $\bm{\theta}$ are obtained anew for every layer,
so an updated state has no ties to former states.
The adaptive character of ADAPT-VQE should lead to implementations with shallower circuits~\cite{grimsley2019adaptive,Feniou2023}.

A crucial point for the optimal convergence towards the target state is the choice of excitation operators $A_k$. These are predefined in an operator pool, prior to the start of the simulation.
Since our interest lies in the nuclear shell model, with a Hamiltonian of the form in~Eq.~(\ref{eq:smham}), we use a pool of two-body fermionic excitation operators 
\begin{equation}\label{eq:ferm_pool}
    T_{rs}^{pq} = i (a_p^{\dag} a_q^{\dag} a_r a_s - a_r^{\dag} a_s^{\dag} a_p a_q),
\end{equation}
where $p,q,r$ and $s$ are single-particle labels with quantum numbers $n,l,j,m$ and $t_z$. 
The same operator may be selected more than once throughout the iterative process, but not on consecutive iterations. 
We apply symmetry considerations when building the Slater determinant basis for the nuclear ground state, 
and only consider excitation operators which conserve the total angular momentum and isospin projection $M$ and $T_z$. This iterative procedure continues until convergence, defined when all the gradient norms in~Eq.~(\ref{eq:gradient}) vanish and/or when the energy 
is close enough to a known solution from, for instance,
classical diagonalization benchmarks. While one could consider more complex operators, involving triple or quadruple particle-hole excitations~\cite{stetcu,papenbrock}, our simulations
indicate that, for the wide set of nuclei studied in this work, 
full shell-model correlations can be captured
at the two-body level with a commensurate number of ansatz layers, of at most a few hundred.

\subsection*{Circuit design strategy}

The main aim of this paper is to determine 
the optimal architecture of quantum circuits that can implement a
nuclear shell-model VQE. 
We explore all the necessary stages of a VQE, from the encoding to 
the energy measurement in the Methods section. 
Ultimately, the circuit design strategy that we propose provides

an approximation-free implementation of ADAPT-VQE,
 in a one-to-one correspondence with the method~\cite{grimsley2019adaptive,tang2021qubit}. 
Having access
to the circuit structure across the full VQE minimization process,
including energy measurements, is a key step forward in discussing the 
scalability of nuclear shell-model simulations in quantum devices,
and it is particularly critical to estimate 
the necessary resources for nuclear shell-model simulations 
with a real quantum advantage, that 
is, in isotopes or regions of the chart where current classical 
devices cannot be employed.

We benchmark our circuit implementation
with circuit-free ADAPT-VQE simulations~\cite{romeroquantum}. 
The latter implement the full algorithm using regular matrix calculus, expressing statevectors, Hamiltonians and pool operators as sparse matrices in the Fock basis. 
With the circuit for the ansatz built and optimized, we simulate the energy measurement protocol, to test the circuits for the changes of basis needed to extract energies in an actual quantum computer.

The state preparation protocol is the most resource-intensive part 
of the algorithm and we provide indications of the resource costs in the Simulations
subsection. We can also quantify and optimize the scaling of the energy 
measurements. 
The nuclear shell-model Hamiltonian in Eq.~(\ref{eq:smham}) consists of one and two-body operators, which can be expressed in terms of Pauli strings (see 
the Methods section).
The one-body part of the Hamiltonian is diagonal and can be measured directly.
We divide the two-body part in three different kinds of terms, depending
on the number of repeated indices. 
Table~\ref{tab:meas} lists the number of circuits needed to measure the expectation value of each part of the Hamiltonian for the \emph{p}, \emph{sd} and \emph{pf} shells. 
Our design strategy indicates that $100$ circuits should suffice to compute any isotope in the $p$ shell and semi-magic nuclei in the $sd$ shell. Open-shell isotopes require a factor of $4-6$ more circuits than their semi-magic counterparts in a given shell.  

In a quantum computer implementation, an energy calculation will be affected by statistical errors. Across a whole ADAPT-VQE simulation, the 
 total number of circuits to be measured for each layer will be
 the product of three terms, $N_s\times N_{tot}\times N_{fc}$.
 The number of shots, $N_s$, is of statistical nature and,
 as discussed in the Methods section in the context of Eq.~(\ref{eq:nshots}),
 it will be sensitive to error mitigation schemes. 
 $N_{tot}$ is the number of different energy measurement circuits.
 We estimate this number and show the results in Table~\ref{tab:meas}.
 Finally, $N_{fc}$ is the number of function calls from the classical 
 optimizer, which we analyze in the Supplementary Information.

\begin{table}[t]
\begin{center}
\begin{tabular}{c|c|c|c|c}
 \text{shell} & $N_{qb}$ & $N_\text{h}$ & $N_{\text{hh}}$ & $N_{tot}$
 \\ \hline
 \multirow{2}{*}{\centering \emph{p}}  & 6 & 2 & 10 (9) & 13 (12)
 \\ \cline{2-5}
                                  & 12 & 4 & 109 (44) & 114 (49)
\\ \hline                
 \multirow{2}{*}{\centering \emph{sd}}  & 12 & 8 & 203 (86) & 212 (95)
 \\ \cline{2-5}
                                  & 24 & 16 & 1389 (518) & 1406 (535)
\\ \hline                
 \multirow{2}{*}{\centering \emph{pf}}  & 20 & 20 & 1507 (570) & 1528 (591)
 \\ \cline{2-5}
                                  & 40 & 40 & 10572 (3459) & 10613 (3500)
\end{tabular}
\end{center}
\caption{{\bf Number of circuits needed to measure the expectation value of the nuclear shell-model for the \emph{p}, \emph{sd} and \emph{pf} shells.}
 $N_{qb}$ indicates the number of qubits for only neutrons or protons (top row for each shell) or both nucleon types (bottom). $N_{\text{h}}$ and $N_{\text{hh}}$ are the number of single-  and double-hopping  terms in the Hamiltonian (related to $h_{ijki}$ and $h_{ijkl}$, respectively), defining the number of circuits needed to measure these parts. The last column lists the total number of circuits, $N_{\text{h}}+N_{\text{hh}}+1$, accounting also for the single circuit needed to measure $\langle n_i\rangle $ and $\langle h_{ijij}^{(l)}\rangle $. The values in parenthesis correspond to the minimum number of groups containing $h_{ijkl}$ terms that commute with each other and thus can be measured with the same circuit. 
}
\label{tab:meas}
\end{table}

\subsection*{Simulations}\label{sec:simulation}

\begin{table}[t]
\begin{center}
\begin{tabular}{c|c|c|c|c|c|c}
 \text{shell} & $N_{qb}$  & $N_{\text{SD}}$  &  \text{nucleus} & $N_{{\rm layers}}$ & $\varepsilon_E$ bound 
 & $N_{{\rm C}}$ (bound)\\ \hline
 \multirow{5}{*}{\centering \emph{p}}  &  6 &  5 & $^6{}$Be    &  2  & $10^{-8}$ &42 (80)\rule{0pt}{2.0ex} \\ \cline{2-7}
           & \multirow{4}{*}{\centering 12} &  10 &  $^6{}$Li   &  9 &  $10^{-7}$&92 (176) \rule{0pt}{2.0ex} \\
                                           &&  53 & $^8{}$Be    &  48 & $10^{-7}$ & 68 (176) \\
                                           &&  51 & $^{10}{}$Be &  48 & $10^{-7}$ & 62 (176) \\
                                          &&  21  & $^{13}{}$C  &  19 & $10^{-7}$& 77 (176) \\\hline
 \multirow{7}{*}{\centering \emph{sd}} & 
  \multirow{4}{*}{\centering 12} &  14 & $^{18}{}$O  &  5  &  $10^{-6}$&99 (176)\rule{0pt}{2.0ex}  \\
                                 &&  37 & $^{19}{}$O  &  32 &  $10^{-6}$&85 (176) \\
                                 &&  81 & $^{20}{}$O  &  70 &  $10^{-6}$&98 (176) \\
                                 &&  142 & $^{22}{}$O  & 117 &  $10^{-6}$&93 (176) \\ \cline{2-7}
  & \multirow{3}{*}{\centering 24} &  640 & $^{20}{}$Ne & 167 &  $2\times 10^{-2}$&137 (368)\rule{0pt}{2.0ex}  \\
                                 &&  4206 & $^{22}{}$Ne & 236 &  $2\times 10^{-2}$&137 (368)  \\
                                 &&  7562 & $^{24}{}$Ne & 345 &  $2\times 10^{-2}$&138 (368)  \\ \hline
 \multirow{5}{*}{\centering \emph{pf}} & 
 \multirow{5}{*}{\centering 20}         &  30  & $^{42}{}$Ca &  9  &  $10^{-8}$&116 (304)\rule{0pt}{2.0ex} \\
                                       &&  565  & $^{44}{}$Ca & 132 &  $10^{-2}$&153 (304)  \\
                                      &&  3952   & $^{46}{}$Ca & 124 &  $10^{-2}$&139 (304)  \\
                                      &&  12022   & $^{48}{}$Ca & 101 &  $10^{-2}$&137 (304)  \\
                                     &&  17276   & $^{50}{}$Ca & 221 &  $10^{-2}$&130 (304) 
\end{tabular}
\end{center}
\caption{ {\bf Ansatz and circuit depth for a given energy bound.}
Number of ansatz layers ($N_{{\rm layers}}$) and relative-error ($\varepsilon_E$) upper bounds for the ground-state energy of all nuclei simulated in this work, organized according to their configuration space ($p$, $sd$, and $pf$ shells), number of qubits $N_{qb}$, and of many-body configurations (Slater determinants) $N_{SD}$. The last column reports the average number of CNOT gates per layer $N_{\rm C}$ together with its upper bound, $16(N_{qb}-2)$ (see Methods). For nuclei with $N_{{\rm layers}}>100$, the average only accounts for the first $100$ layers. }
\label{tab:numlayers}
\end{table}

 The systems we explore include nuclei across different shells, 
 with even and odd numbers of protons and neutrons (see panel (a) of Fig.~\ref{fig:master}). 
We find that circuit-free and circuit-full simulations employing
the same parameter minimization algorithm agree to numerical accuracy.
 
We estimate the required depth of a circuit by imposing bounds on the 
relative error of the ground-state energy, 
$\displaystyle \varepsilon_E= \frac{\lvert  E-E_{\textrm{SM}}  \lvert}{E_{\textrm{SM}}}$, 
where $E_{\textrm{SM}}$ is the corresponding classical shell-model diagonalization
result. 
Table~\ref{tab:numlayers} lists the number of ADAPT-VQE layers needed in an ansatz state 
to achieve a given value of $\varepsilon_E$ for 
a series of nuclei across the $p$, $sd$ and $pf$ shells. 
All energies tend to converge to the benchmark values, albeit with different rates.
Semi-magic nuclei close to the closed shell typically converge rapidly, with less than $10$ ADAPT-VQE layers. In contrast, 
the most costly nuclei simulated in this work, neon isotopes, require a few hundred ADAPT-VQE layers to reach a ground-state energy error of $2\%$. 
Nonetheless, we stress that the optimizations do not get stuck in barren plateaus.
A key advantage of our circuit design strategy is that it allows
us to quantify the associated quantum circuit resources. We take the
number of CNOT gates required in the state preparation, 
$N_\text{CNOT}$, as a quantitative indicator of 
circuit resources.

\begin{figure}[t]
     \centering
     \includegraphics[width=\linewidth]{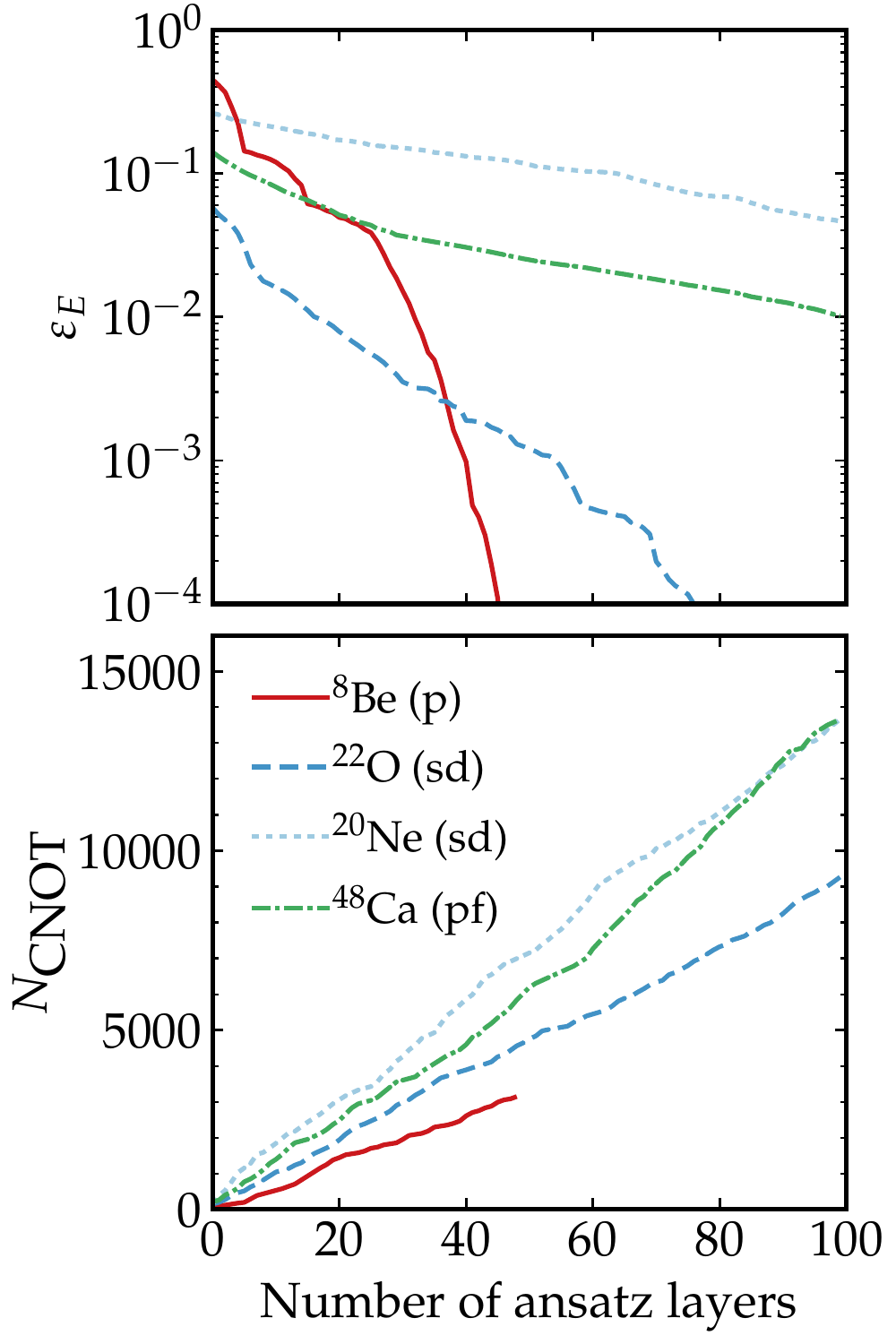}
     \caption{
     {\bf Energy relative error and circuit complexity as a function of ADAPT-VQE layers.}
     Evolution of the relative error for the ground-state energy, $\varepsilon_{\rm E}$, (top panel) and number of CNOT gates in the ansatz circuit (bottom) as a function of the number of ansatz layers for simulations of $^{8}$Be, $^{22}$O, $^{20}$Ne and $^{48}$Ca. As the algorithm adaptively iterates, errors decay exponentially while the number of CNOT gates increases linearly or polynomially.}
     \label{fig:evolution-nuclei}
\end{figure}

Figure~\ref{fig:evolution-nuclei} shows the evolution of $\varepsilon_E$ (top panel) and $N_\text{CNOT}$ (bottom) as a function of the number of ADAPT-VQE layers for four representative isotopes across different nuclear shells. Simulations for all nuclei show that $\varepsilon_E$ decreases exponentially as the number of layers in the ansatz increases, while the number of CNOT gates grows linearly or polynomially. 

This number depends on the particular operators chosen by the ADAPT-VQE minimization, but it is at most $16\,(N_{qb}-1)$ per ansatz layer (see Methods section). In contrast, the average number of CNOT gates per ansatz layer found by ADAPT-VQE simulations is

roughly half of the corresponding upper bounds, see Table~\ref{tab:numlayers}.
As an example, finding the ground-state energy of $^{22}$O with an error of few percent, requires about $20$ ansatz layers and $\approx 2000$
CNOT gates. We provide more details for all the nuclei studied in this work
in the Supplementary Information. 
Figure~\ref{fig:evolution-nuclei} and Table~\ref{tab:numlayers} demonstrate that ADAPT-VQE converges exponentially as the number of layers, or equivalently CNOT gates, is increased.

\begin{figure}[t!]
     \centering
     \includegraphics[width=\linewidth]{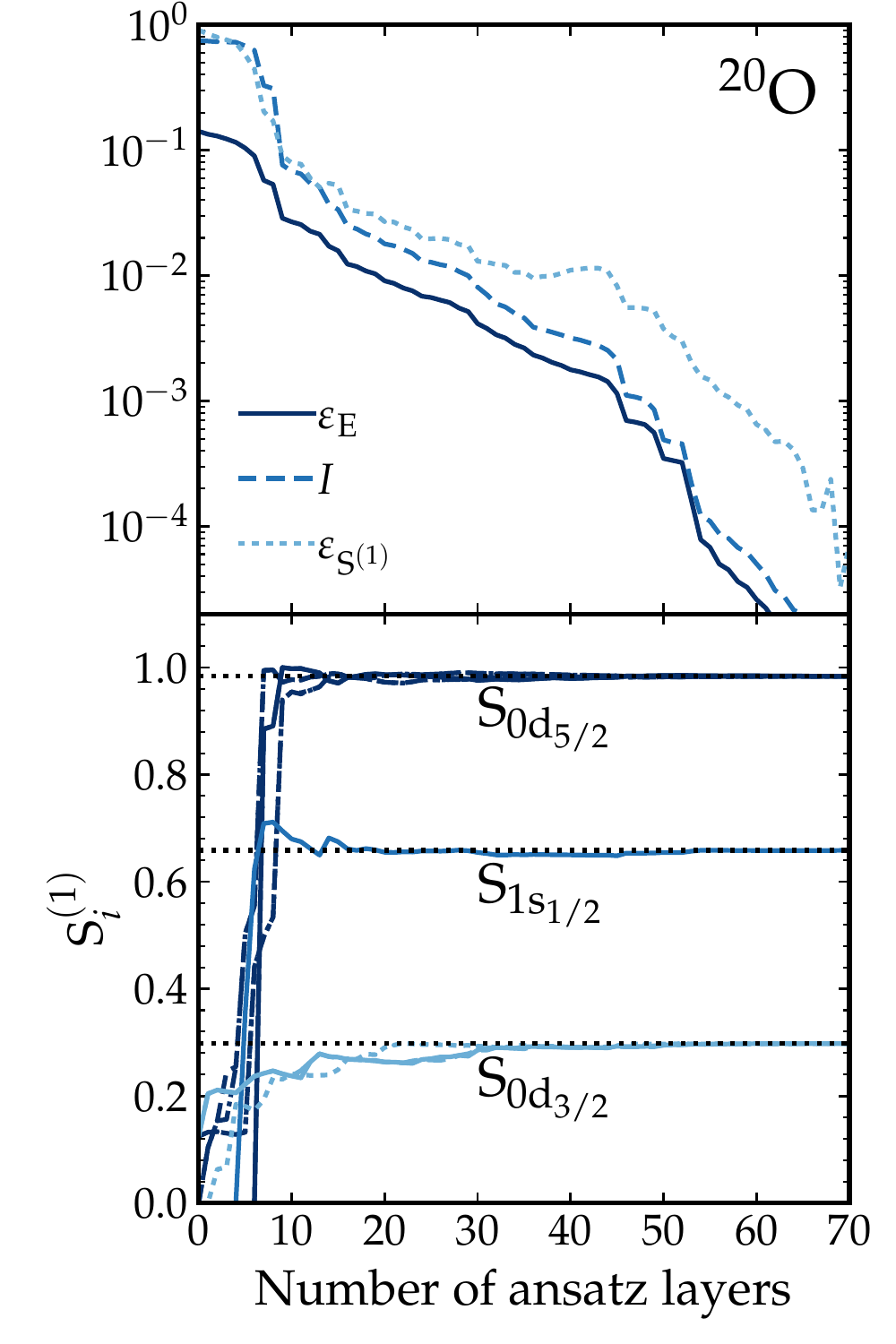}
     \caption{
     {\bf Quality of the wavefunction and entanglement entropy as a function of ADAPT-VQE layers.}
    Evolution of the relative error for the ground-state energy, $\varepsilon_{\rm E}$, the infidelity, $I$, and the average relative error of single-orbital entropies, $\varepsilon_{S^{(1)}}$ for $^{20}$O as a function of the number of ansatz layers (top panel). Evolution of $S_i^{(1)}$ for the same nucleus and $i$ orbitals $0d_{3/2}$, $1s_{1/2}$ and $0d_{5/2}$, where the dotted lines indicate the entropies for the exact solution (bottom panel). The maximum $S_k^{(1)}$ is $1$, very close to the value of the $0d_{5/2}$ orbitals.}
     \label{fig:evolution-FS}
\end{figure}

Our results are either commensurate or competitive  compared to
previous estimates of circuit depth based on UCC-VQE
on the $p$ shell and
on two oxygen isotopes on the $sd$ shell~\cite{stetcu,papenbrock}. 
For $^8$Be, Stetcu \emph{et al.} require $112$ variational parameters to reach $\varepsilon_E \approx 1 \%$ even after including triple and quadruple excitation operators~\cite{stetcu}. Our implementation of ADAPT-VQE, with two-body excitation operators only, requires $48$ parameters to reach $\varepsilon_E=10^{-7}$. 
In $^{22}$O, the UCC-VQE ansatz leads to $\varepsilon_E \approx 3 \%$ with $35$
parameters~\cite{stetcu}, 
whereas Fig.~\ref{fig:evolution-FS} indicates that ADAPT-VQE
reaches a similar level of accuracy with about $20$ layers. For $^6$Li, we find that $9$ layers suffice to get a converged result up to $10^{-7}$,
in contrast to the observations of Ref.~\cite{papenbrock}, where an alternative
ADAPT-VQE implementation reaches only $\varepsilon_E \approx 10^{-3}$. 
A difference between previous implementations and our work is that 
we let our classical minimizer reach bottom precision at each ADAPT-VQE
layer, whereas Kiss \emph{et al.} employ $10$ minimization steps per layer (with the SPSA
optimiser)~\cite{papenbrock}. 
Moreover, UCC-VQE shell-model implementations have so far relied on Hartree-Fock 
reference states, 
which may not be optimal starting points for VQEs~\cite{romeroquantum,Stetcu2023}. 
Either way, it appears that ADAPT-VQE shell-model simulations 
outperform their UCC-VQE counterparts in terms of layers, 
an observation that is in line with findings in quantum 
chemistry~\cite{haidar2022open}. 
We note, however, that an unbiased comparison of quantum hardware efficiency between different methods requires
a one-to-one quantification of the resources in each approach, including explicitly energy measurement overheads. 

ADAPT-VQE predicts the ground-state energy of the nucleus, but one also has access to the nuclear wavefunction $|\psi (\bm{\theta})\rangle$,
although reconstructing it from quantum hardware may  require costly quantum tomography. 
One can quantify the quality with respect to a given benchmark wavefunction, $|\psi_\text{b} \rangle$, by employing the infidelity 
$I= 1-|\langle \psi_\text{b}|\psi (\bm{\theta}) \rangle|^2$.
We  take the classical shell model as a benchmark, and the better the level of agreement between
both wavefunctions, the closer $I$ is to $0$. 
We also use the single-orbital entanglement
entropy, 
$S_{i} = 
-(1-\gamma_{i}) \log_2 (1-\gamma_{i}) -\gamma_{i} \log_2 \gamma_{i} $, with
$\gamma_{i}= \langle \psi (\bm{\theta}) \lvert a^\dagger_i a_i \lvert \psi (\bm{\theta}) \rangle$
, bound between $0$ and $1$,
to evaluate the importance of quantum correlations in the ansatz~\cite{Gigena2015,Robin2021,johnson2023proton,bulgac2023measures,Pazy2023,Bulgac2022b}.

These two indicators provide quantitative complementary information on the quality of the wavefunction and
the variational process.

Focusing on the test case example of 
$^{20}$O, the top panel of Fig.~\ref{fig:evolution-FS} 
shows the infidelity $I$ of the ground state with respect to the shell-model wavefunction (dashed line). 
The panel also shows the average of relative errors 
of each single-particle state entanglement entropy, 
$\varepsilon_{S^{(1)}}=\frac{1}{N_{qb}}\sum_i \varepsilon_{S_i^{(1)}}$ (dotted line). These two quantities follow closely $\varepsilon_E$ 
along the iterative process.
We observe a few sudden drops in the relative error for the energy, which correlate with similar drops in $I$ and $\varepsilon_{S^{(1)}}$. This indicates that, at certain points in the optimization, ADAPT-VQE entangles parts of the nucleus relatively faster than others. 
Overall, the curves suggest that the ADAPT-VQE ansatz captures efficiently the entanglement structure of the many-body wavefunction.
A more extensive analysis of the infidelity is provided in the Supplementary
Information. 
The bottom panel of Fig.~\ref{fig:evolution-FS} provides a closer inspection to the entanglement structure of this nucleus. Based on previous
studies~\cite{Robin2021,stetcu,johnson2023proton}, we expect nuclear-structure features to correlate with single-particle states entanglement properties.
The panel shows the quantum simulated single-orbital entropies of the $12$ single-particle
states as a function of the number of ansatz layers, compared to the classical shell-model entropies (horizontal 
dotted lines). We clearly distinguish the emergence of three subshells
in the entropy. 
The most entangled qubits are those in the lowest-energy orbital, $0{d_{5/2}}$, reaching almost the maximal value.  
These are followed by the $1{s_{1/2}}$ and the $0{d_{3/2}}$ states, which 
are correspondingly less entangled (and occupied). 
The entropies saturate to the
shell-model value relatively quickly,
within about 20 layers. We take this as an indication that ADAPT-VQE 
captures early on the most important correlations of the nucleus, which are 
subsequently refined by the variational process. 
\section*{Discussion}\label{sec:conclusions}

In this work, we provide a detailed framework for a quantum hardware implementation of ADAPT-VQE tailored to nuclear shell-model calculations. 
The algorithm requires as many qubits as the number of single-particle states, a relatively small number ($\approx 50$) even for valence spaces demanding currently unavailable classical computational resources. 
We benchmark our results with calculations using a circuit-free, regular matrix implementation of the algorithm.

Our simulations do not become stuck in local minima or barren plateaus.
We find that the majority of the resources in the quantum circuit are dedicated 
to the construction of the parametrized ansatz wave function. 
Each additional parameter in the ansatz increases the circuit depth linearly
with the number of qubits. 
In contrast, the preparation of the reference state and the implementation of the basis changes to measure Hamiltonian expectation values are comparatively small parts of the total circuit depth. We quantify (see Methods) the number of 
circuits needed to measure energies in the different isotopes. 
Our proposed energy-measuring circuits are not substantially deeper than the
corresponding circuit encoding the wave function. 

\begin{figure}[b!]
     \centering
     \includegraphics[width=\linewidth]{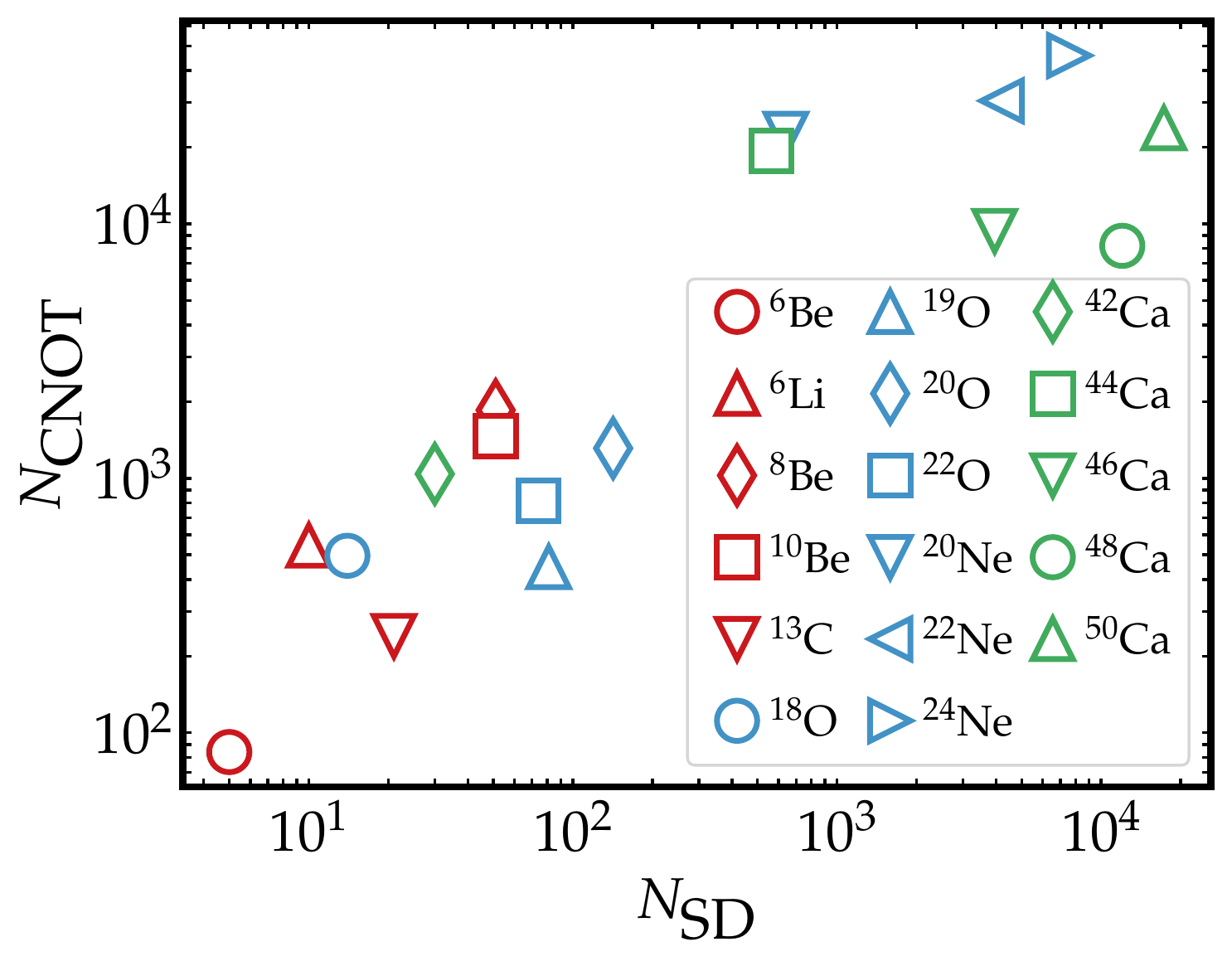}
     \caption{
     {\bf Correlation between number of CNOTs and Slater determinants.}
Total number of CNOTs $N_{\textrm{CNOT}}$ needed to obtain a ground-state relative energy error of $2\%$ as a function of the number of Slater determinants $N_{\textrm{SD}}$ in the many-body basis for all nuclei considered in this work. The observed trend does not indicate a quantum exponential advantage over classical methods.}
     \label{fig:cnotslater2}
\end{figure}

We calculate the ground state of selected nuclei in the \emph{p-}, \emph{sd-} and \emph{pf-}shell valence spaces, using up to 24 qubits.  
For all these systems, our simulations indicate that the relative error in the ground-state energy and the infidelity decrease exponentially as
the number of layers in the ansatz increases (see Supplementary Information).
While the number of parameters needed to reach a certain precision depends on the nucleus, our results indicate that at most $150$ CNOT gates per ADAPT-VQE layer 
are necessary to get ground-state energies accurate at the percent level. This suggests that
a circuit implementation of the shell model with ADAPT-VQE may be a 
suitable way forward for quantum computing simulations of nuclei. Nevertheless,
the number of layers and CNOTs shown in Table~\ref{tab:numlayers} do not
demonstrate an exponential quantum advantage~\cite{lee2023evaluating} with respect to the classical computation cost. This is indeed seen more clearly in Fig.~\ref{fig:cnotslater2}, which shows the number of total CNOTs needed to obtain an energy relative error of $2\%$, as a function of the number of Slater determinants for all nuclei studied in this work. Figure~\ref{fig:cnotslater2} indicates that up to nuclear masses $A \simeq 50$ the number of CNOT gates scales roughly as the number of Slater determinants.

Our study opens several potential avenues for further exploration. 
First, 
different fermionic encodings may reduce the number of CNOT gates, which are subject to noise errors that can limit realistic implementations 
in quantum devices. A preliminary analysis using the
Bravyi-Kitaev basis~\cite{Seeley2012} (instead of a JW transformation) suggests a $\approx 10\%$ reduction  in the number of CNOT gates of ADAPT-VQE after $100$ iterations in the $sd$ and $pf$ shells. Other options of fermionic mappings such as Gray code encoding~\cite{di2021improving, siwach2021quantum} should also be explored.
Second, the present work is an ideal testbed for the implementation of quantum information tools for the study of nuclear structure. Our calculated single-particle state entropies reveal the entanglement structure of nuclei, in close analogy to the occupation probabilities of the orbitals obtained in classical diagonalization schemes. Other correlation measures, such as quantum discord~\cite{faba2021correlation,faba2021two, faba3},
 will be the subject of future work. 
Furthermore, one should elucidate more clearly the sharp differences between the UCC and ADAPT ansatz VQEs. On the one hand, the choice of initial states, at the mean-field level~\cite{papenbrock,stetcu} or mixing many-body configurations, may improve the overall performance~\cite{romeroquantum, ph2} of the minimization process. 
On the other, understanding why the ordering in the choice of operators 
is so relevant
may provide  further insights 
into nuclear many-body correlations.
A better understanding on these issues is key to find optimal algorithms and circuit designs for the nuclear shell model that avoid the exponential scaling of resources and can be realistically implemented in NISQ devices. 

We note that there are promising alternative algorithms for nuclear shell-model calculations based on the Lanczos method~\cite{kirby2023exact}.

\section*{Methods}

We simulate circuits for several $p$-, $sd$- and $pf$-shell nuclei using the statevector simulator {\sc qibo}~\cite{Efthymiou_2022}, together with the {\sc qibojit} package, which harnesses multi-core parallelization based on JIT (just-in-time) compilation and the 
{\sc numba}  compiler~\cite{lam15}. 
{\sc qibo} has been found to be specially efficient when compared to other simulators for similar fermionic quantum-circuit simulations~\cite{perezobiol_2022}. 
At each layer, we execute the quantum circuit to extract a statevector $|\psi_n\rangle$ of dimension $2^{N_{qb}}$.
This extraction is limited by classical computer
resources, which in turn provide stringent mass limits for our classical
circuit simulations. For instance, simulating open-shell nuclei in the $pf$ shell valence space,
requires state-vectors with $2^{40}$ complex coefficients, 
demanding $8$~TB of memory in single-precision format. 
When dealing with $20$ or more qubits, we use GPUs and the {\sc cupy} compiler~\cite{okuta17cupy} to accelerate computations.

\begin{table}[t]
\begin{center}
\begin{tabular}{c|c|c}
  & \text{Fermion Operators} & \text{Qubit Operators} \\ \hline
$n_p$   &  $a_p^{\dag}a_p$  &  $\dfrac12 (1-Z_p)\Tstrut\Bstrut$   \\ \hline 
$h_{pqrs}$  &  $\begin{aligned}a_p^{\dag}a_q^{\dag}a_r a_s \\+ a_r^{\dag}a_s^{\dag}a_p a_q\end{aligned}$  &
$\begin{aligned} \dfrac18 P_{rs}^{pq}\,
(
&-X_p X_q X_r X_s + X_p X_q Y_r Y_s
\\&- X_p Y_q X_r Y_s- X_p Y_q Y_r X_s
\\&- Y_p Y_q Y_r Y_s + Y_p Y_q X_r X_s
\\&- Y_p X_q Y_r X_s - Y_p X_q X_r Y_s 
)\end{aligned}\rule{0pt}{8.5ex}$ 
\\ \hline
$T_{rs}^{pq}$ & $\begin{aligned}i(a_p^{\dag}a_q^{\dag}a_r a_s \\- a_r^{\dag}a_s^{\dag}a_p a_q)\end{aligned}$
&
$\begin{aligned}\dfrac18 P_{rs}^{pq}\,
(
&-X_p Y_q Y_r Y_s - Y_p X_q Y_r Y_s 
\\&+ Y_p Y_q X_r Y_s  + Y_p Y_q Y_r X_s
\\&+Y_p X_q X_r X_s + X_p Y_q X_r X_s
\\& - X_p X_q Y_r X_s - X_p X_q X_r Y_s 
)\end{aligned}\rule{0pt}{8.5ex}$
\\ \hline
$h_{pq}$   & $a_p^{\dag}a_q + a_q^{\dag}a_p$  &
$ \dfrac12\left(\prod_{n=p+1}^{q-1}Z_n\right)\left(X_p X_q+Y_q Y_p\right)\Tstrut\Bstrut$
\\ \hline
$T_{pq}$  & $i(a_p^{\dag}a_q - a_q^{\dag} a_p)$  &
$\dfrac12\left(\prod_{n=p+1}^{q-1}Z_n\right)\left(Y_p X_q-X_q Y_p\right)\Tstrut\Bstrut$
\end{tabular}
\end{center}
\caption{{\bf Jordan-Wigner transformation for the main operators appearing in the Hamiltonian and in our ADAPT-VQE operator pool.} Indices run over $p<q$ and $r<s$, assuming that all are different. If two indices are repeated, then $h_{pqpr}=-n_p h_{qr}$ and
$T_{pq}^{pr}=n_p T_{qr}$, with $q<r$. We note that $h_{pqpq}=-2n_p n_q$ and $T_{pq}^{pq}=0$.
}
\label{tab:ops}
\end{table}

Next, we describe the five different stages~\cite{Tilly2022} of our VQE circuit design strategy.

\subsection*{Mapping}

We consider the JW mapping~\cite{jw,Seeley2012}, which transforms nucleonic creation and annihilation operators as
\begin{equation}
a_i^{\dag} = \left(\prod_{k=0}^{i-1}Z_k\right)\sigma^-_i,~~
a_i = \left(\prod_{k=0}^{i-1} Z_k\right)\sigma^+_i,
\end{equation}
where $\sigma^\pm_j=\frac12(X_j\pm i Y_j)$ and $X_j$, $Y_j$, $Z_j$ are the usual Pauli matrices applied to qubit $j$. Using these relations we can express any fermionic operator in terms of Pauli strings. Table~\ref{tab:ops} lists the expressions for the two types of (self-adjoint) terms appearing in the nuclear shell-model Hamiltonian $H_{\rm{eff}}$ in~Eq.~(\ref{eq:smham}). 
We use an auxiliary operator
\begin{equation}\label{eq:aux_ope}
P_{rs}^{pq} \equiv \left(\prod_{m=p+1, m\notin [r,s]}^{q-1}Z_m\right)\left(\prod_{n=r+1,n\notin [p,q]}^{s-1}Z_n\right).
\end{equation}
Table~\ref{tab:ops} also indicates the JW transformation
for the pool operators $T_{rs}^{pq}$, and for single-excitation operators which appear when indices are repeated in either $h_{pqrs}$ or $T_{rs}^{pq}$.
In this context, the most important features of an operator are the numbers and lengths of the Pauli strings they contain. These ultimately determine the efficiency in the circuit implementation of ADAPT-VQE. 
The two operators $h_{pqrs}$ and $T_{rs}^{pq}$ contain $8$ Pauli strings, each of length
$L_{pqrs}=n_2+n_4-n_1-n_3+2$,
where $n_1$, $n_2$, $n_3$ and $n_4$ are the indices $p$, $q$, $r$ and $s$ sorted in ascending order. For example, if $(p,q,r,s)=(2,8,5,7)$, then $(n_1,n_2,n_3,n_4)=(2,5,7,8)$ and $L_{2857}=6$. If two indices are repeated, the expressions simplify to $h_{pqpr}$ and $T_{pq}^{pr}$, as indicated in Table~\ref{tab:ops}. These consist of two Pauli strings of length $L_{pqr}^{(1)}=r-q+1$ and two other strings of length $L_{pqr}^{(2)}=r-q+2$.

\begin{figure}[tb]
    \centering
    \includegraphics[width=\linewidth]{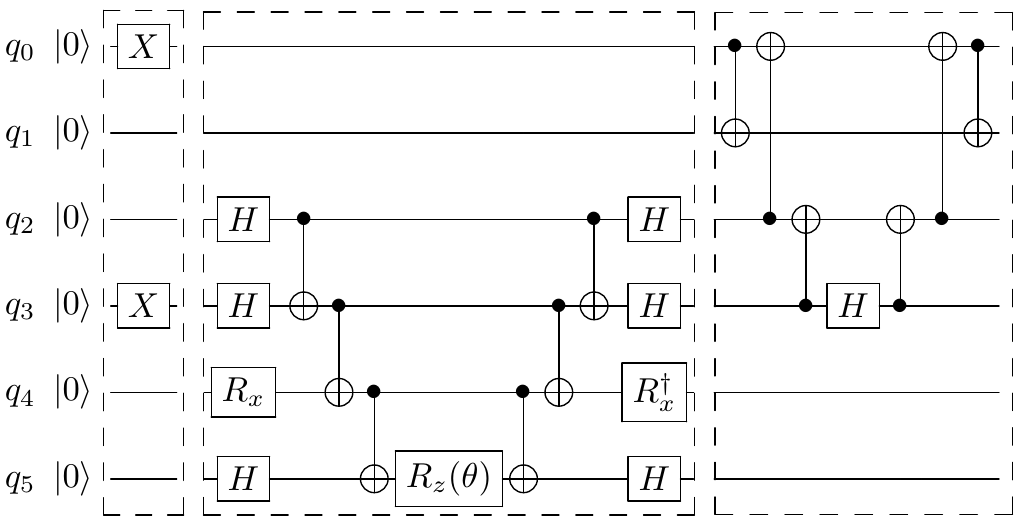}
    \caption{{\bf Examples of main circuit blocks, separated by dashed boxes, in ADAPT-VQE for the simulation of $^6$Be.} Left: preparation of the reference state defined in~Eq.~(\ref{eq:refbe6f}) and~Eq.~(\ref{eq:refbe6q}). Middle: implementation of $e^{-i\frac{\theta}{2} X_2X_3Y_4Z_5}$ using the CNOT staircase algorithm, one out of the many unitaries in the variational part of ADAPT-VQE. 
    Right: circuit of the basis change $M_{0123}$ needed to diagonalize $h_{0123}$. The subcircuit in qubits $q_2$ and $q_3$ containing two CNOTs and a Hadamard gate $H$ corresponds to the basis change $M_{23}$.}
    \label{fig:circuit_blocks}
\end{figure}
\subsection*{Initial state preparation}
To provide a minimal starting point to the simulations, 
we choose the lowest-energy Slater determinant as a reference state. 
Under the JW mapping, Slater determinants are mapped to the computational basis by flipping the qubits corresponding to the occupied orbitals using $X$ gates. 
Considering for example the case of $^6$Be, an isotope in the $p$ shell (panel (b) of Fig.~\ref{fig:master}) and for our interaction of choice, the lowest-energy Slater determinant is 
\begin{equation}
   |0,3\rangle = a_0^{\dag} a_3^{\dag}\,|\rm{vac}\rangle,
    \label{eq:refbe6f}
\end{equation}
where $|\rm{vac}\rangle$ is the vacuum state with no particles in the valence space. 
After a JW mapping, the state is translated into the computational basis as 
\begin{equation}
    |100100\rangle = X_0 X_3|000000\rangle.
\label{eq:refbe6q}
\end{equation}
The leftmost block of Fig.~\ref{fig:circuit_blocks} shows the corresponding circuit.

This choice of initial state preparation is minimal in terms of circuit resources: it has
unit depth independently of the number of orbitals in the valence space and it does not involve any two-qubit gates. 
For a given valence neutron 
and proton number, $N_\text{CI}$ and $Z_\text{CI}$, finding the lowest energy Slater determinant requires at most
 
$N_\text{SD}$ operations.
This task can be performed relatively quickly in a classical computer,
and is a one-off pre-processing overhead that we do not incorporate
in the circuit resources discussed below.

\begin{figure*}[tb]
    \centering
\includegraphics[width=\linewidth]{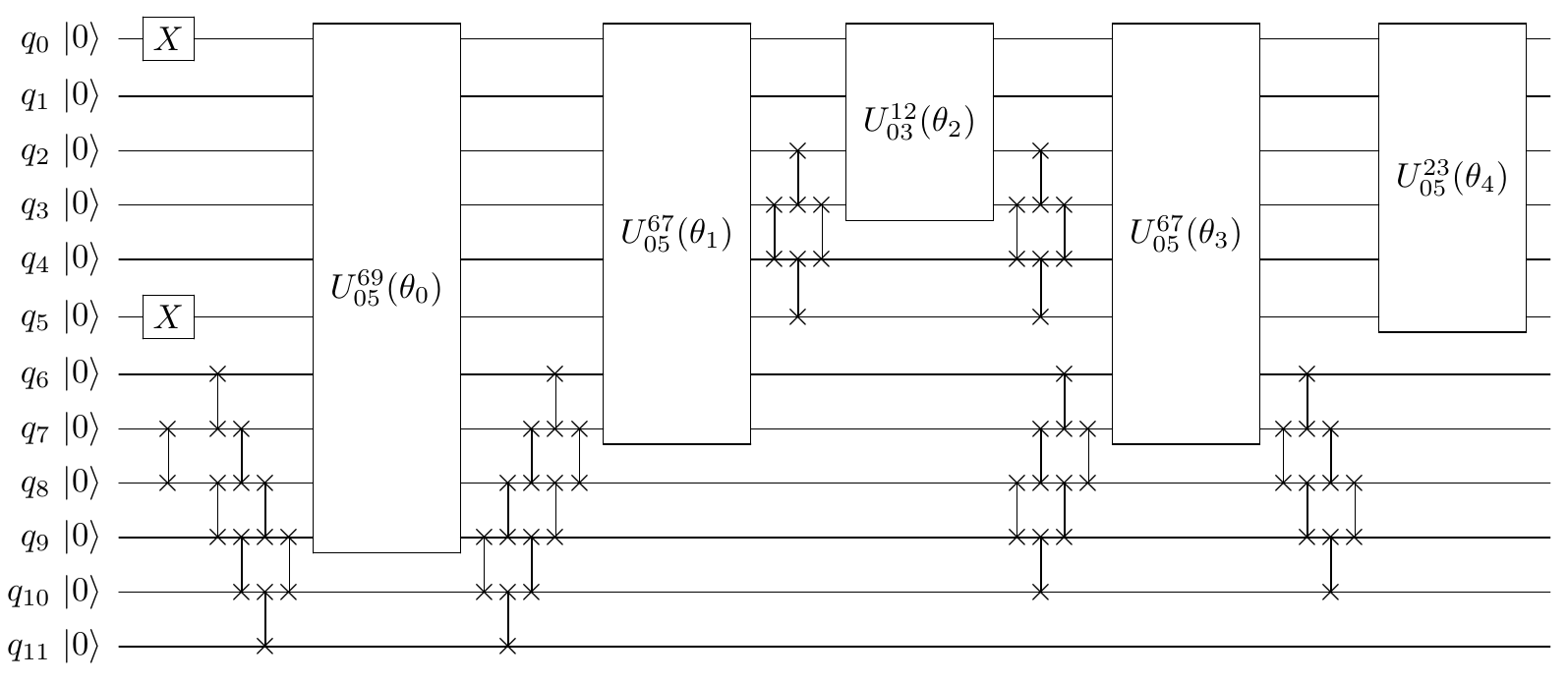}
\caption{{\bf Circuit to prepare the $^{18}$O ground state.} $X$ gates prepare the reference state and FSWAP gates change the basis so that pool-operator exponentials act on adjacent qubits. Multiqubit gates in boxes are defined as $U_{rs}^{pq}(\theta)\equiv e^{i\theta T_{rs}^{pq}}$ and $\theta_0=-0.157263$, $\theta_1=-0.437238$, $\theta_2=0.604663$, $\theta_3=0.214431$, $\theta_4=-0.785469$.
}
\label{fig:circuit20}
\end{figure*}
\subsection*{Variational optimization}
\label{sec:variational}

The variational ansatz is parametrized as in~Eq.~(\ref{eq:adaptwf1}), with pool operators $A_k=T_{rs}^{pq}$ given in Table~\ref{tab:ops} after the JW transformation. 
We convert the pool operators $T_{rs}^{pq}$ to Pauli strings using the OpenFermion package~\cite{openfermion}, and for the circuits for the unitaries $e^{i\theta T_{rs}^{pq}}$ we follow the staircase algorithm of Fig.~\ref{fig:circuit_blocks}. In the simulated circuits we only use single-qubit and CNOT gates.

All Pauli strings in these sums commute with each other, so each term in $T_{rs}^{pq}$ can be exponentiated separately and there is no need for a Trotter-Suzuki approximation. This results in the expression
\begin{equation}
\begin{split}
e^{i\theta T_{rs}^{pq}}
=&
e^{-i\theta' P_{rs}^{pq} X_p Y_q Y_r Y_s}
e^{-i\theta' P_{rs}^{pq} Y_p X_q Y_r Y_s} \\ \times &
e^{i\theta' P_{rs}^{pq} Y_p Y_q X_r Y_s}
e^{i\theta' P_{rs}^{pq} Y_p Y_q Y_r X_s}\\\times & 
e^{i\theta' P_{rs}^{pq} Y_p X_q X_r X_s}
e^{i\theta' P_{rs}^{pq} X_p Y_q X_r X_s}\\ \times&
e^{-i\theta' P_{rs}^{pq} X_p X_q Y_r X_s}
e^{-i\theta' P_{rs}^{pq} X_p X_q X_r Y_s },
\end{split}
\end{equation}
with $\theta'=\theta/8$ and $P_{rs}^{pq}$ given in~Eq.~(\ref{eq:aux_ope}).
The exponential of a single Pauli string is particularly easy to implement with the staircase algorithm~\cite{sawaya20}. If the Pauli string contains only $Z$ matrices, the circuit contains two cascades of CNOTs and a $Z$ rotation, $R_z(\theta)\equiv e^{-i \frac{\theta}{2}Z}$, with $-\frac{\theta}{2}$ the coefficient multiplying the Pauli string.
If the product contains an $X$ or $Y$ matrix, we apply a basis change in the corresponding qubit, namely $X=HZH$ and $Y=R_x^\dagger Z R_x$, where $H$ is the Hadamard gate and $R_x$ the rotation $e^{-i \frac{\pi}{4}X}$.
Figure~\ref{fig:circuit_blocks} (middle) illustrates the procedure for the example implementation of $e^{-i\frac{\theta}{2} X_2X_3Y_4X_5}$.
If $e^{i\theta T_{rs}^{pq}}$ acts on non-adjacent qubits, we implement a change of basis through fermionic SWAP (FSWAP) gates, so that only CNOTs applied to contiguous qubits are needed. The FSWAP exchanges states while maintaining the correct parity,
\begin{equation}
    \text{FSWAP}=1+a_i^\dagger a_j+a_j^\dagger a_i-a_i^\dagger a_i-a_j^\dagger a_j.
\end{equation}
Using the staircase protocol, each parametrized layer $e^{i\theta T_{rs}^{pq}}$ requires $16\,(L_{pqrs}-1)$ CNOT gates, where $L_{pqrs}$ is the average length of the Pauli strings in the operator.
$L_{pqrs}$ is bounded by the number of qubits $N_{qb}$, implying that the maximum number of CNOTs per ansatz layer is $16\,(N_{qb}-1)$ and that the depth per layer grows linearly with the number of single-particle states in the valence space. If qubits are linearly connected in hardware and non-adjacent qubit states are brought together with FSWAPs, the depth per layer has a total linear overhead. The precise overhead size depends on how qubits are arranged and connected to each other. However, it is bounded by $4(N_{qb}-4)$.

Let us provide an example illustrating the simplicity of the ADAPT-VQE circuit implementation. 
Obtaining the ground-state energy of simple nuclei only demands a few operators. As shown in Results, 
ADAPT-VQE simulations for 
$^{18}$O converge to an energy accuracy better than $10^{-6}$ with a five-layer ansatz, reading 
\begin{equation}
|\psi_{ ^{18}{\rm O}}\rangle = e^{i\theta_4 T^{05}_{23}}e^{i\theta_3 T^{05}_{9\,10}}e^{i\theta_2 T^{05}_{14}}e^{i\theta_1 T^{05}_{67}}e^{i\theta_0 T^{05}_{8\,11}}X_0X_5|0\rangle^{\otimes 12} .
\nonumber
\end{equation}
Figure~\ref{fig:circuit20} shows the full circuit assuming one-dimensional connectivity between qubits, and gives the parameter values.
Our algorithm includes the multiqubit operators $e^{i\theta T_{rs}^{pq}}$
involving CNOT gates acting on non-adjacent qubits when these are laid out in a one-dimensional array. We manipulate these operators to include only local two-qubit gates through a series of FSWAPs.
\subsection*{Measurement}\label{sec:measurement}
Once the ADAPT-VQE ansatz $|\psi_n \rangle$ is prepared in the quantum circuit at a given layer $n$, we measure the energy with the expectation value $\langle \psi_n|H_{\rm{eff}}|\psi_n\rangle$. 
To this end, we build a series of circuits that implement a change of basis to 
diagonalize separately each term of the Hamiltonian. 
The number of terms in the shell-model Hamiltonian scales with the number of qubits as $O(N_{qb}^4)$, but we find a much milder
scaling of the circuit number with $N_{qb}$.

One-body (number) operators $n_i$ are diagonal and can be  measured directly,
\begin{equation}
\langle \psi_n|n_i|\psi_n\rangle = \frac12 \langle \psi_n|1-Z_i|\psi_n\rangle = p_{1}^{(i)},
\end{equation}
where $p_{1}^{(i)} $, the probability of measuring $``1"$ in qubit $i$, can be extracted by measuring multiple times that qubit. Since all one-body operators commute with each other, we can measure all of them simultaneously.
The two-body part of the Hamiltonian $h_{ijkl}$ can be divided into three kinds of terms depending on whether indices ($i$, $j$, $k$, $l$) are two, three, or four different integers.
Local terms $h_{ijij}$ are the product of two number operators $n_i$ and $n_j$ and they can be measured simultaneously,
\begin{equation}
\langle \psi_n|h_{ijij}|\psi_n\rangle  = -2\langle \psi_n|n_i n_j|\psi_n\rangle = -2 p_{11}^{(ij)},
\end{equation}
with $p_{11}^{(ij)} $ the probability to measure $``1"$ in qubits $i$ and $j$.
The non-diagonal parts of $h_{ijik}$ and  $h_{ijkl}$
swap two states in the subspaces of qubits ($i$, $j$, $k$)
and ($i$, $j$, $k$, $l$), respectively. 
These operators can be disentangled through series of CNOT gates and reduced to an $X$ gate acting on a single qubit. The Pauli matrix $X$ is then diagonalized with a Hadamard gate, $X = H Z H$. In turn, we diagonalize  $h_{ijik}$ and  $h_{ijkl}$ using $M_{jk}\equiv\,CX_{kj}\,H_k\,CX_{kj}$ and $M_{ijkl}\equiv\, CX_{ij}CX_{ki}CX_{lk} H_l CX_{lk} CX_{ki} CX_{ij}$, where $CX_{ij}$ represents a CNOT
 gate with control qubit $i$ and target qubit $j$.
 The right block of Fig.~\ref{fig:circuit_blocks} illustrates the corresponding circuit implementation. 
After diagonalization, assuming contiguous indices, the expectation values read
\begin{equation}\label{eq:ndham1}
\begin{split}
\langle \psi_n|h_{ijik}|\psi_n\rangle = &
p_{101}^{(ijk)}-p_{110}^{(ijk)},
\end{split}
\end{equation}
and 
\begin{equation}\label{eq:ndham2}
\begin{split}
\langle \psi_n|h_{ijkl}|\psi_n\rangle =&
p_{1100}^{(ijkl)} -  p_{0011}^{(ijkl)},
\end{split}
\end{equation}
with $ p_{r_1 \cdots r_k}^{(q_1 \dots q_k)}$ being the probabilities of measuring results $r_1$ to $r_k$ in qubits $q_1$ to $q_k$ in the statevector where the basis changes have been applied. We refer to the Supplementary Information for a detailed derivation 
of~Eq.~(\ref{eq:ndham1}) and 
Eq.~(\ref{eq:ndham2}).

 The changes of basis needed for measurements add, for any nucleus, an overhead of zero, two or six two-qubit gates depending on the Hamiltonian term measured. This represents a small fraction of the circuit depth and a constant scaling with the number of single-particle states in the valence space. 
We discuss in the Supplementary Information
details regarding to the number of different
measurement circuits required to measure the
energy as well as the gradients of 
Eq.~(\ref{eq:gradient}).

\subsection*{Error mitigation}

Finally, expectation values of the Hamiltonian computed using the algorithm described above are subject to statistical errors and quantum noise. The former scale as the inverse of the number of shots, $\sigma_E \propto \frac{1}{\sqrt{N_{s}}}$. In other words, given a target error in the energy accuracy $\varepsilon_{\langle H\rangle}$, the number of necessary shots scales as
\begin{equation}
    N_s\propto\frac{1}{\varepsilon_{\langle H\rangle}^2}.
    \label{eq:nshots}
\end{equation}
The specific factor may be estimated simulating the measurement protocol. 
A straightforward and robust strategy to mitigate errors for ADAPT-VQE shell-model simulations
is to use symmetry considerations and discard measurements that do not yield results consistent with the Fock basis of the simulated nucleus. Since the JW mapping identifies Fock and computational states, this amounts to excluding all states with different number of measured $``1$''s than nucleons in the valence space.
Likewise, one should also ignore states with measured $``1$''s distributed in a set of qubits corresponding to a different angular momentum or isospin than the simulated nucleus. This protocol should be particularly effective in mitigating single bit-flip errors, which effectively create or destroy nucleons, as well as multiple bit-flip errors which do not preserve either nucleon number, angular momentum or isospin. These simple but robust 
strategies may be key in future implementations of this method on NISQ devices. 

\section*{Data availability}
The data that support the findings of this study are available within the paper and its Supplementary Information. 
Any additional information
is available from the corresponding authors upon request.

\bibliography{sample}

\begin{thebibliography}{87}%
\makeatletter
\providecommand \@ifxundefined [1]{%
 \@ifx{#1\undefined}
}%
\providecommand \@ifnum [1]{%
 \ifnum #1\expandafter \@firstoftwo
 \else \expandafter \@secondoftwo
 \fi
}%
\providecommand \@ifx [1]{%
 \ifx #1\expandafter \@firstoftwo
 \else \expandafter \@secondoftwo
 \fi
}%
\providecommand \natexlab [1]{#1}%
\providecommand \enquote  [1]{``#1''}%
\providecommand \bibnamefont  [1]{#1}%
\providecommand \bibfnamefont [1]{#1}%
\providecommand \citenamefont [1]{#1}%
\providecommand \href@noop [0]{\@secondoftwo}%
\providecommand \href [0]{\begingroup \@sanitize@url \@href}%
\providecommand \@href[1]{\@@startlink{#1}\@@href}%
\providecommand \@@href[1]{\endgroup#1\@@endlink}%
\providecommand \@sanitize@url [0]{\catcode `\\12\catcode `\$12\catcode
  `\&12\catcode `\#12\catcode `\^12\catcode `\_12\catcode `\%12\relax}%
\providecommand \@@startlink[1]{}%
\providecommand \@@endlink[0]{}%
\providecommand \url  [0]{\begingroup\@sanitize@url \@url }%
\providecommand \@url [1]{\endgroup\@href {#1}{\urlprefix }}%
\providecommand \urlprefix  [0]{URL }%
\providecommand \Eprint [0]{\href }%
\providecommand \doibase [0]{http://dx.doi.org/}%
\providecommand \selectlanguage [0]{\@gobble}%
\providecommand \bibinfo  [0]{\@secondoftwo}%
\providecommand \bibfield  [0]{\@secondoftwo}%
\providecommand \translation [1]{[#1]}%
\providecommand \BibitemOpen [0]{}%
\providecommand \bibitemStop [0]{}%
\providecommand \bibitemNoStop [0]{.\EOS\space}%
\providecommand \EOS [0]{\spacefactor3000\relax}%
\providecommand \BibitemShut  [1]{\csname bibitem#1\endcsname}%
\let\auto@bib@innerbib\@empty
\bibitem [{\citenamefont {Taniuchi}\ \emph {et~al.}(2019)\citenamefont
  {Taniuchi} \emph {et~al.}}]{Taniuchi:2019pen}%
  \BibitemOpen
  \bibfield  {author} {\bibinfo {author} {\bibfnamefont {R.}~\bibnamefont
  {Taniuchi}} \emph {et~al.},\ }\href {\doibase 10.1038/s41586-019-1155-x}
  {\bibfield  {journal} {\bibinfo  {journal} {Nature}\ }\textbf {\bibinfo
  {volume} {569}},\ \bibinfo {pages} {53} (\bibinfo {year} {2019})},\ \Eprint
  {http://arxiv.org/abs/arxiv:1912.05978} {arXiv:arxiv:1912.05978 [nucl-ex]}
  \BibitemShut {NoStop}%
\bibitem [{\citenamefont {Butler}\ \emph {et~al.}(2019)\citenamefont {Butler}
  \emph {et~al.}}]{Butler:2019qox}%
  \BibitemOpen
  \bibfield  {author} {\bibinfo {author} {\bibfnamefont {P.~A.}\ \bibnamefont
  {Butler}} \emph {et~al.},\ }\href {\doibase 10.1038/s41467-019-10494-5}
  {\bibfield  {journal} {\bibinfo  {journal} {Nat. Commun.}\ }\textbf {\bibinfo
  {volume} {10}},\ \bibinfo {pages} {2473} (\bibinfo {year} {2019})},\ \Eprint
  {http://arxiv.org/abs/arxiv:2003.10147} {arXiv:arxiv:2003.10147 [nucl-ex]}
  \BibitemShut {NoStop}%
\bibitem [{\citenamefont {Tsunoda}\ \emph {et~al.}(2020)\citenamefont
  {Tsunoda}, \citenamefont {Otsuka}, \citenamefont {Takayanagi}, \citenamefont
  {Shimizu}, \citenamefont {Suzuki}, \citenamefont {Utsuno}, \citenamefont
  {Yoshida},\ and\ \citenamefont {Ueno}}]{Tsunoda:2020gpt}%
  \BibitemOpen
  \bibfield  {author} {\bibinfo {author} {\bibfnamefont {N.}~\bibnamefont
  {Tsunoda}}, \bibinfo {author} {\bibfnamefont {T.}~\bibnamefont {Otsuka}},
  \bibinfo {author} {\bibfnamefont {K.}~\bibnamefont {Takayanagi}}, \bibinfo
  {author} {\bibfnamefont {N.}~\bibnamefont {Shimizu}}, \bibinfo {author}
  {\bibfnamefont {T.}~\bibnamefont {Suzuki}}, \bibinfo {author} {\bibfnamefont
  {Y.}~\bibnamefont {Utsuno}}, \bibinfo {author} {\bibfnamefont
  {S.}~\bibnamefont {Yoshida}}, \ and\ \bibinfo {author} {\bibfnamefont
  {H.}~\bibnamefont {Ueno}},\ }\href {\doibase 10.1038/s41586-020-2848-x}
  {\bibfield  {journal} {\bibinfo  {journal} {Nature}\ }\textbf {\bibinfo
  {volume} {587}},\ \bibinfo {pages} {66} (\bibinfo {year} {2020})}\BibitemShut
  {NoStop}%
\bibitem [{\citenamefont {Schmidt}\ \emph {et~al.}(2020)\citenamefont {Schmidt}
  \emph {et~al.}}]{CLAS:2020mom}%
  \BibitemOpen
  \bibfield  {author} {\bibinfo {author} {\bibfnamefont {A.}~\bibnamefont
  {Schmidt}} \emph {et~al.} (\bibinfo {collaboration} {CLAS}),\ }\href
  {\doibase 10.1038/s41586-020-2021-6} {\bibfield  {journal} {\bibinfo
  {journal} {Nature}\ }\textbf {\bibinfo {volume} {578}},\ \bibinfo {pages}
  {540} (\bibinfo {year} {2020})},\ \Eprint
  {http://arxiv.org/abs/arxiv:2004.11221} {arXiv:arxiv:2004.11221 [nucl-ex]}
  \BibitemShut {NoStop}%
\bibitem [{\citenamefont {Mukha}\ \emph {et~al.}(2006)\citenamefont {Mukha}
  \emph {et~al.}}]{Mukha06}%
  \BibitemOpen
  \bibfield  {author} {\bibinfo {author} {\bibfnamefont {I.}~\bibnamefont
  {Mukha}} \emph {et~al.},\ }\href {\doibase 10.1038/nature04453} {\bibfield
  {journal} {\bibinfo  {journal} {Nature}\ }\textbf {\bibinfo {volume} {439}},\
  \bibinfo {pages} {298} (\bibinfo {year} {2006})}\BibitemShut {NoStop}%
\bibitem [{\citenamefont {Hinke}\ \emph {et~al.}(2012)\citenamefont {Hinke}
  \emph {et~al.}}]{Hinke12}%
  \BibitemOpen
  \bibfield  {author} {\bibinfo {author} {\bibfnamefont {C.~B.}\ \bibnamefont
  {Hinke}} \emph {et~al.},\ }\href {\doibase 10.1038/nature11116} {\bibfield
  {journal} {\bibinfo  {journal} {Nature}\ }\textbf {\bibinfo {volume} {486}},\
  \bibinfo {pages} {341} (\bibinfo {year} {2012})}\BibitemShut {NoStop}%
\bibitem [{\citenamefont {Walz}\ \emph {et~al.}(2015)\citenamefont {Walz},
  \citenamefont {Scheit}, \citenamefont {Pietralla}, \citenamefont {Aumann},
  \citenamefont {Lefol},\ and\ \citenamefont {Ponomarev}}]{Walz15}%
  \BibitemOpen
  \bibfield  {author} {\bibinfo {author} {\bibfnamefont {C.}~\bibnamefont
  {Walz}}, \bibinfo {author} {\bibfnamefont {H.}~\bibnamefont {Scheit}},
  \bibinfo {author} {\bibfnamefont {N.}~\bibnamefont {Pietralla}}, \bibinfo
  {author} {\bibfnamefont {T.}~\bibnamefont {Aumann}}, \bibinfo {author}
  {\bibfnamefont {R.}~\bibnamefont {Lefol}}, \ and\ \bibinfo {author}
  {\bibfnamefont {V.~Y.}\ \bibnamefont {Ponomarev}},\ }\href {\doibase
  10.1038/nature15543} {\bibfield  {journal} {\bibinfo  {journal} {Nature}\
  }\textbf {\bibinfo {volume} {526}},\ \bibinfo {pages} {406} (\bibinfo {year}
  {2015})}\BibitemShut {NoStop}%
\bibitem [{\citenamefont {Cowan}\ \emph {et~al.}(2021)\citenamefont {Cowan},
  \citenamefont {Sneden}, \citenamefont {Lawler}, \citenamefont {Aprahamian},
  \citenamefont {Wiescher}, \citenamefont {Langanke}, \citenamefont
  {Mart\'\i{}nez-Pinedo},\ and\ \citenamefont {Thielemann}}]{Cowan:2019pkx}%
  \BibitemOpen
  \bibfield  {author} {\bibinfo {author} {\bibfnamefont {J.~J.}\ \bibnamefont
  {Cowan}}, \bibinfo {author} {\bibfnamefont {C.}~\bibnamefont {Sneden}},
  \bibinfo {author} {\bibfnamefont {J.~E.}\ \bibnamefont {Lawler}}, \bibinfo
  {author} {\bibfnamefont {A.}~\bibnamefont {Aprahamian}}, \bibinfo {author}
  {\bibfnamefont {M.}~\bibnamefont {Wiescher}}, \bibinfo {author}
  {\bibfnamefont {K.}~\bibnamefont {Langanke}}, \bibinfo {author}
  {\bibfnamefont {G.}~\bibnamefont {Mart\'\i{}nez-Pinedo}}, \ and\ \bibinfo
  {author} {\bibfnamefont {F.-K.}\ \bibnamefont {Thielemann}},\ }\href
  {\doibase 10.1103/RevModPhys.93.015002} {\bibfield  {journal} {\bibinfo
  {journal} {Rev. Mod. Phys.}\ }\textbf {\bibinfo {volume} {93}},\ \bibinfo
  {pages} {15002} (\bibinfo {year} {2021})},\ \Eprint
  {http://arxiv.org/abs/arxiv:1901.01410} {arXiv:arxiv:1901.01410
  [astro-ph.HE]} \BibitemShut {NoStop}%
\bibitem [{\citenamefont {Aalbers}\ \emph {et~al.}(2023)\citenamefont {Aalbers}
  \emph {et~al.}}]{Aalbers:2022dzr}%
  \BibitemOpen
  \bibfield  {author} {\bibinfo {author} {\bibfnamefont {J.}~\bibnamefont
  {Aalbers}} \emph {et~al.},\ }\href {\doibase 10.1088/1361-6471/ac841a}
  {\bibfield  {journal} {\bibinfo  {journal} {J. Phys. G}\ }\textbf {\bibinfo
  {volume} {50}},\ \bibinfo {pages} {013001} (\bibinfo {year} {2023})},\
  \Eprint {http://arxiv.org/abs/arxiv:2203.02309} {arXiv:arxiv:2203.02309
  [physics.ins-det]} \BibitemShut {NoStop}%
\bibitem [{\citenamefont {Engel}\ \emph {et~al.}(2013)\citenamefont {Engel},
  \citenamefont {Ramsey-Musolf},\ and\ \citenamefont {van
  Kolck}}]{Engel:2013lsa}%
  \BibitemOpen
  \bibfield  {author} {\bibinfo {author} {\bibfnamefont {J.}~\bibnamefont
  {Engel}}, \bibinfo {author} {\bibfnamefont {M.~J.}\ \bibnamefont
  {Ramsey-Musolf}}, \ and\ \bibinfo {author} {\bibfnamefont {U.}~\bibnamefont
  {van Kolck}},\ }\href {\doibase 10.1016/j.ppnp.2013.03.003} {\bibfield
  {journal} {\bibinfo  {journal} {Prog. Part. Nucl. Phys.}\ }\textbf {\bibinfo
  {volume} {71}},\ \bibinfo {pages} {21} (\bibinfo {year} {2013})},\ \Eprint
  {http://arxiv.org/abs/arxiv:1303.2371} {arXiv:arxiv:1303.2371 [nucl-th]}
  \BibitemShut {NoStop}%
\bibitem [{\citenamefont {Avignone}\ \emph {et~al.}(2008)\citenamefont
  {Avignone}, \citenamefont {Elliott},\ and\ \citenamefont
  {Engel}}]{Avignone:2007fu}%
  \BibitemOpen
  \bibfield  {author} {\bibinfo {author} {\bibfnamefont {F.~T.}\ \bibnamefont
  {Avignone}, \bibfnamefont {III}}, \bibinfo {author} {\bibfnamefont {S.~R.}\
  \bibnamefont {Elliott}}, \ and\ \bibinfo {author} {\bibfnamefont
  {J.}~\bibnamefont {Engel}},\ }\href {\doibase 10.1103/RevModPhys.80.481}
  {\bibfield  {journal} {\bibinfo  {journal} {Rev. Mod. Phys.}\ }\textbf
  {\bibinfo {volume} {80}},\ \bibinfo {pages} {481} (\bibinfo {year} {2008})},\
  \Eprint {http://arxiv.org/abs/arxiv:0708.1033} {arXiv:arxiv:0708.1033
  [nucl-ex]} \BibitemShut {NoStop}%
\bibitem [{\citenamefont {Mayer}(1949)}]{mayerII}%
  \BibitemOpen
  \bibfield  {author} {\bibinfo {author} {\bibfnamefont {M.~G.}\ \bibnamefont
  {Mayer}},\ }\href {\doibase https://doi.org/10.1103/PhysRev.75.1969}
  {\bibfield  {journal} {\bibinfo  {journal} {Phys. Rev.}\ }\textbf {\bibinfo
  {volume} {75}},\ \bibinfo {pages} {1969} (\bibinfo {year}
  {1949})}\BibitemShut {NoStop}%
\bibitem [{\citenamefont {Haxel}\ \emph {et~al.}(1949)\citenamefont {Haxel},
  \citenamefont {Jensen},\ and\ \citenamefont {Suess}}]{Haxel:1949fjd}%
  \BibitemOpen
  \bibfield  {author} {\bibinfo {author} {\bibfnamefont {O.}~\bibnamefont
  {Haxel}}, \bibinfo {author} {\bibfnamefont {J.~H.~D.}\ \bibnamefont
  {Jensen}}, \ and\ \bibinfo {author} {\bibfnamefont {H.~E.}\ \bibnamefont
  {Suess}},\ }\href {\doibase 10.1103/PhysRev.75.1766.2} {\bibfield  {journal}
  {\bibinfo  {journal} {Phys. Rev.}\ }\textbf {\bibinfo {volume} {75}},\
  \bibinfo {pages} {1766} (\bibinfo {year} {1949})}\BibitemShut {NoStop}%
\bibitem [{\citenamefont {Brown}\ and\ \citenamefont
  {Wildenthal}(1988)}]{brown1988status}%
  \BibitemOpen
  \bibfield  {author} {\bibinfo {author} {\bibfnamefont {B.~A.}\ \bibnamefont
  {Brown}}\ and\ \bibinfo {author} {\bibfnamefont {B.}~\bibnamefont
  {Wildenthal}},\ }\href {\doibase 10.1146/annurev.ns.38.120188.000333}
  {\bibfield  {journal} {\bibinfo  {journal} {Annu. Rev. Nucl. Part. Sci.}\
  }\textbf {\bibinfo {volume} {38}},\ \bibinfo {pages} {29} (\bibinfo {year}
  {1988})}\BibitemShut {NoStop}%
\bibitem [{\citenamefont {Caurier}\ \emph {et~al.}(2005)\citenamefont
  {Caurier}, \citenamefont {Martinez-Pinedo}, \citenamefont {Nowacki},
  \citenamefont {Poves},\ and\ \citenamefont {Zuker}}]{Caurier:2004gf}%
  \BibitemOpen
  \bibfield  {author} {\bibinfo {author} {\bibfnamefont {E.}~\bibnamefont
  {Caurier}}, \bibinfo {author} {\bibfnamefont {G.}~\bibnamefont
  {Martinez-Pinedo}}, \bibinfo {author} {\bibfnamefont {F.}~\bibnamefont
  {Nowacki}}, \bibinfo {author} {\bibfnamefont {A.}~\bibnamefont {Poves}}, \
  and\ \bibinfo {author} {\bibfnamefont {A.~P.}\ \bibnamefont {Zuker}},\ }\href
  {\doibase 10.1103/RevModPhys.77.427} {\bibfield  {journal} {\bibinfo
  {journal} {Rev. Mod. Phys.}\ }\textbf {\bibinfo {volume} {77}},\ \bibinfo
  {pages} {427} (\bibinfo {year} {2005})},\ \Eprint
  {http://arxiv.org/abs/arxiv:nucl-th/0402046} {arXiv:arxiv:nucl-th/0402046}
  \BibitemShut {NoStop}%
\bibitem [{\citenamefont {Otsuka}\ \emph {et~al.}(2020)\citenamefont {Otsuka},
  \citenamefont {Gade}, \citenamefont {Sorlin}, \citenamefont {Suzuki},\ and\
  \citenamefont {Utsuno}}]{otsuka2020evolution}%
  \BibitemOpen
  \bibfield  {author} {\bibinfo {author} {\bibfnamefont {T.}~\bibnamefont
  {Otsuka}}, \bibinfo {author} {\bibfnamefont {A.}~\bibnamefont {Gade}},
  \bibinfo {author} {\bibfnamefont {O.}~\bibnamefont {Sorlin}}, \bibinfo
  {author} {\bibfnamefont {T.}~\bibnamefont {Suzuki}}, \ and\ \bibinfo {author}
  {\bibfnamefont {Y.}~\bibnamefont {Utsuno}},\ }\href {\doibase
  10.1103/RevModPhys.92.015002} {\bibfield  {journal} {\bibinfo  {journal}
  {Rev. Mod. Phys.}\ }\textbf {\bibinfo {volume} {92}},\ \bibinfo {pages}
  {015002} (\bibinfo {year} {2020})},\ \Eprint
  {http://arxiv.org/abs/arxiv:1805.06501} {arxiv:1805.06501} \BibitemShut
  {NoStop}%
\bibitem [{\citenamefont {Stroberg}\ \emph {et~al.}(2019)\citenamefont
  {Stroberg}, \citenamefont {Bogner}, \citenamefont {Hergert},\ and\
  \citenamefont {Holt}}]{Stroberg:2019mxo}%
  \BibitemOpen
  \bibfield  {author} {\bibinfo {author} {\bibfnamefont {S.~R.}\ \bibnamefont
  {Stroberg}}, \bibinfo {author} {\bibfnamefont {S.~K.}\ \bibnamefont
  {Bogner}}, \bibinfo {author} {\bibfnamefont {H.}~\bibnamefont {Hergert}}, \
  and\ \bibinfo {author} {\bibfnamefont {J.~D.}\ \bibnamefont {Holt}},\ }\href
  {\doibase 10.1146/annurev-nucl-101917-021120} {\bibfield  {journal} {\bibinfo
   {journal} {Annu. Rev. Nucl. Part. Sci.}\ }\textbf {\bibinfo {volume} {69}},\
  \bibinfo {pages} {307} (\bibinfo {year} {2019})},\ \Eprint
  {http://arxiv.org/abs/arxiv:1902.06154} {arXiv:arxiv:1902.06154 [nucl-th]}
  \BibitemShut {NoStop}%
\bibitem [{\citenamefont {Arute}\ \emph {et~al.}(2019)\citenamefont {Arute},
  \citenamefont {Arya}, \citenamefont {Babbush}, \citenamefont {Bacon},
  \citenamefont {Bardin}, \citenamefont {Barends}, \citenamefont {Biswas},
  \citenamefont {Boixo}, \citenamefont {Brandao}, \citenamefont {Buell} \emph
  {et~al.}}]{arute2019quantum}%
  \BibitemOpen
  \bibfield  {author} {\bibinfo {author} {\bibfnamefont {F.}~\bibnamefont
  {Arute}}, \bibinfo {author} {\bibfnamefont {K.}~\bibnamefont {Arya}},
  \bibinfo {author} {\bibfnamefont {R.}~\bibnamefont {Babbush}}, \bibinfo
  {author} {\bibfnamefont {D.}~\bibnamefont {Bacon}}, \bibinfo {author}
  {\bibfnamefont {J.~C.}\ \bibnamefont {Bardin}}, \bibinfo {author}
  {\bibfnamefont {R.}~\bibnamefont {Barends}}, \bibinfo {author} {\bibfnamefont
  {R.}~\bibnamefont {Biswas}}, \bibinfo {author} {\bibfnamefont
  {S.}~\bibnamefont {Boixo}}, \bibinfo {author} {\bibfnamefont {F.~G.}\
  \bibnamefont {Brandao}}, \bibinfo {author} {\bibfnamefont {D.~A.}\
  \bibnamefont {Buell}},  \emph {et~al.},\ }\href {\doibase
  10.1038/s41586-019-1666-5} {\bibfield  {journal} {\bibinfo  {journal}
  {Nature}\ }\textbf {\bibinfo {volume} {574}},\ \bibinfo {pages} {505}
  (\bibinfo {year} {2019})}\BibitemShut {NoStop}%
\bibitem [{\citenamefont {Preskill}(2018)}]{preskill2018quantum}%
  \BibitemOpen
  \bibfield  {author} {\bibinfo {author} {\bibfnamefont {J.}~\bibnamefont
  {Preskill}},\ }\href {\doibase 10.22331/q-2018-08-06-79} {\bibfield
  {journal} {\bibinfo  {journal} {Quantum}\ }\textbf {\bibinfo {volume} {2}},\
  \bibinfo {pages} {79} (\bibinfo {year} {2018})},\ \Eprint
  {http://arxiv.org/abs/arxiv:1801.00862} {arxiv:1801.00862} \BibitemShut
  {NoStop}%
\bibitem [{\citenamefont {Peruzzo}\ \emph {et~al.}(2014)\citenamefont
  {Peruzzo}, \citenamefont {McClean}, \citenamefont {Shadbolt}, \citenamefont
  {Yung}, \citenamefont {Zhou}, \citenamefont {Love}, \citenamefont
  {Aspuru-Guzik},\ and\ \citenamefont {O’brien}}]{peruzzo2014variational}%
  \BibitemOpen
  \bibfield  {author} {\bibinfo {author} {\bibfnamefont {A.}~\bibnamefont
  {Peruzzo}}, \bibinfo {author} {\bibfnamefont {J.}~\bibnamefont {McClean}},
  \bibinfo {author} {\bibfnamefont {P.}~\bibnamefont {Shadbolt}}, \bibinfo
  {author} {\bibfnamefont {M.-H.}\ \bibnamefont {Yung}}, \bibinfo {author}
  {\bibfnamefont {X.-Q.}\ \bibnamefont {Zhou}}, \bibinfo {author}
  {\bibfnamefont {P.~J.}\ \bibnamefont {Love}}, \bibinfo {author}
  {\bibfnamefont {A.}~\bibnamefont {Aspuru-Guzik}}, \ and\ \bibinfo {author}
  {\bibfnamefont {J.~L.}\ \bibnamefont {O’brien}},\ }\href {\doibase
  10.1038/ncomms5213} {\bibfield  {journal} {\bibinfo  {journal} {Nat.
  Commun.}\ }\textbf {\bibinfo {volume} {5}},\ \bibinfo {pages} {1} (\bibinfo
  {year} {2014})},\ \Eprint {http://arxiv.org/abs/arxiv:1304.3061}
  {arxiv:1304.3061} \BibitemShut {NoStop}%
\bibitem [{\citenamefont {McClean}\ \emph {et~al.}(2016)\citenamefont
  {McClean}, \citenamefont {Romero}, \citenamefont {Babbush},\ and\
  \citenamefont {Aspuru-Guzik}}]{mcclean2016theory}%
  \BibitemOpen
  \bibfield  {author} {\bibinfo {author} {\bibfnamefont {J.~R.}\ \bibnamefont
  {McClean}}, \bibinfo {author} {\bibfnamefont {J.}~\bibnamefont {Romero}},
  \bibinfo {author} {\bibfnamefont {R.}~\bibnamefont {Babbush}}, \ and\
  \bibinfo {author} {\bibfnamefont {A.}~\bibnamefont {Aspuru-Guzik}},\ }\href
  {\doibase 10.1088/1367-2630/18/2/023023} {\bibfield  {journal} {\bibinfo
  {journal} {New J. Phys.}\ }\textbf {\bibinfo {volume} {18}},\ \bibinfo
  {pages} {023023} (\bibinfo {year} {2016})},\ \Eprint
  {http://arxiv.org/abs/arxiv:1509.04279} {arxiv:1509.04279} \BibitemShut
  {NoStop}%
\bibitem [{\citenamefont {Bharti}\ \emph {et~al.}(2022)\citenamefont {Bharti},
  \citenamefont {Cervera-Lierta}, \citenamefont {Kyaw}, \citenamefont {Haug},
  \citenamefont {Alperin-Lea}, \citenamefont {Anand}, \citenamefont {Degroote},
  \citenamefont {Heimonen}, \citenamefont {Kottmann}, \citenamefont {Menke}
  \emph {et~al.}}]{bharti2022noisy}%
  \BibitemOpen
  \bibfield  {author} {\bibinfo {author} {\bibfnamefont {K.}~\bibnamefont
  {Bharti}}, \bibinfo {author} {\bibfnamefont {A.}~\bibnamefont
  {Cervera-Lierta}}, \bibinfo {author} {\bibfnamefont {T.~H.}\ \bibnamefont
  {Kyaw}}, \bibinfo {author} {\bibfnamefont {T.}~\bibnamefont {Haug}}, \bibinfo
  {author} {\bibfnamefont {S.}~\bibnamefont {Alperin-Lea}}, \bibinfo {author}
  {\bibfnamefont {A.}~\bibnamefont {Anand}}, \bibinfo {author} {\bibfnamefont
  {M.}~\bibnamefont {Degroote}}, \bibinfo {author} {\bibfnamefont
  {H.}~\bibnamefont {Heimonen}}, \bibinfo {author} {\bibfnamefont {J.~S.}\
  \bibnamefont {Kottmann}}, \bibinfo {author} {\bibfnamefont {T.}~\bibnamefont
  {Menke}},  \emph {et~al.},\ }\href {\doibase 10.1103/RevModPhys.94.015004}
  {\bibfield  {journal} {\bibinfo  {journal} {Rev. Mod. Phys.}\ }\textbf
  {\bibinfo {volume} {94}},\ \bibinfo {pages} {015004} (\bibinfo {year}
  {2022})},\ \Eprint {http://arxiv.org/abs/arxiv:2101.08448} {arxiv:2101.08448}
  \BibitemShut {NoStop}%
\bibitem [{\citenamefont {Cerezo}\ \emph {et~al.}(2021)\citenamefont {Cerezo},
  \citenamefont {Arrasmith}, \citenamefont {Babbush}, \citenamefont {Benjamin},
  \citenamefont {Endo}, \citenamefont {Fujii}, \citenamefont {McClean},
  \citenamefont {Mitarai}, \citenamefont {Yuan}, \citenamefont {Cincio},\ and\
  \citenamefont {Coles}}]{Cerezo2021}%
  \BibitemOpen
  \bibfield  {author} {\bibinfo {author} {\bibfnamefont {M.}~\bibnamefont
  {Cerezo}}, \bibinfo {author} {\bibfnamefont {A.}~\bibnamefont {Arrasmith}},
  \bibinfo {author} {\bibfnamefont {R.}~\bibnamefont {Babbush}}, \bibinfo
  {author} {\bibfnamefont {S.~C.}\ \bibnamefont {Benjamin}}, \bibinfo {author}
  {\bibfnamefont {S.}~\bibnamefont {Endo}}, \bibinfo {author} {\bibfnamefont
  {K.}~\bibnamefont {Fujii}}, \bibinfo {author} {\bibfnamefont {J.~R.}\
  \bibnamefont {McClean}}, \bibinfo {author} {\bibfnamefont {K.}~\bibnamefont
  {Mitarai}}, \bibinfo {author} {\bibfnamefont {X.}~\bibnamefont {Yuan}},
  \bibinfo {author} {\bibfnamefont {L.}~\bibnamefont {Cincio}}, \ and\ \bibinfo
  {author} {\bibfnamefont {P.~J.}\ \bibnamefont {Coles}},\ }\href {\doibase
  10.1038/s42254-021-00348-9} {\bibfield  {journal} {\bibinfo  {journal} {Nat.
  Rev. Phys.}\ }\textbf {\bibinfo {volume} {3}},\ \bibinfo {pages} {625}
  (\bibinfo {year} {2021})},\ \Eprint {http://arxiv.org/abs/arxiv:2012.09265}
  {arxiv:2012.09265} \BibitemShut {NoStop}%
\bibitem [{\citenamefont {Tilly}\ \emph {et~al.}(2022)\citenamefont {Tilly},
  \citenamefont {Chen}, \citenamefont {Cao}, \citenamefont {Picozzi},
  \citenamefont {Setia}, \citenamefont {Li}, \citenamefont {Grant},
  \citenamefont {Wossnig}, \citenamefont {Rungger}, \citenamefont {Booth} \emph
  {et~al.}}]{Tilly2022}%
  \BibitemOpen
  \bibfield  {author} {\bibinfo {author} {\bibfnamefont {J.}~\bibnamefont
  {Tilly}}, \bibinfo {author} {\bibfnamefont {H.}~\bibnamefont {Chen}},
  \bibinfo {author} {\bibfnamefont {S.}~\bibnamefont {Cao}}, \bibinfo {author}
  {\bibfnamefont {D.}~\bibnamefont {Picozzi}}, \bibinfo {author} {\bibfnamefont
  {K.}~\bibnamefont {Setia}}, \bibinfo {author} {\bibfnamefont
  {Y.}~\bibnamefont {Li}}, \bibinfo {author} {\bibfnamefont {E.}~\bibnamefont
  {Grant}}, \bibinfo {author} {\bibfnamefont {L.}~\bibnamefont {Wossnig}},
  \bibinfo {author} {\bibfnamefont {I.}~\bibnamefont {Rungger}}, \bibinfo
  {author} {\bibfnamefont {G.~H.}\ \bibnamefont {Booth}},  \emph {et~al.},\
  }\href {\doibase https://doi.org/10.1016/j.physrep.2022.08.003} {\bibfield
  {journal} {\bibinfo  {journal} {Phys. Rep.}\ }\textbf {\bibinfo {volume}
  {986}},\ \bibinfo {pages} {1} (\bibinfo {year} {2022})},\ \Eprint
  {http://arxiv.org/abs/arxiv:2111.05176} {arxiv:2111.05176} \BibitemShut
  {NoStop}%
\bibitem [{\citenamefont {Anand}\ \emph {et~al.}(2022)\citenamefont {Anand},
  \citenamefont {Schleich}, \citenamefont {Alperin-Lea}, \citenamefont
  {Jensen}, \citenamefont {Sim}, \citenamefont {D{\'\i}az-Tinoco},
  \citenamefont {Kottmann}, \citenamefont {Degroote}, \citenamefont
  {Izmaylov},\ and\ \citenamefont {Aspuru-Guzik}}]{uccrev}%
  \BibitemOpen
  \bibfield  {author} {\bibinfo {author} {\bibfnamefont {A.}~\bibnamefont
  {Anand}}, \bibinfo {author} {\bibfnamefont {P.}~\bibnamefont {Schleich}},
  \bibinfo {author} {\bibfnamefont {S.}~\bibnamefont {Alperin-Lea}}, \bibinfo
  {author} {\bibfnamefont {P.~W.}\ \bibnamefont {Jensen}}, \bibinfo {author}
  {\bibfnamefont {S.}~\bibnamefont {Sim}}, \bibinfo {author} {\bibfnamefont
  {M.}~\bibnamefont {D{\'\i}az-Tinoco}}, \bibinfo {author} {\bibfnamefont
  {J.~S.}\ \bibnamefont {Kottmann}}, \bibinfo {author} {\bibfnamefont
  {M.}~\bibnamefont {Degroote}}, \bibinfo {author} {\bibfnamefont {A.~F.}\
  \bibnamefont {Izmaylov}}, \ and\ \bibinfo {author} {\bibfnamefont
  {A.}~\bibnamefont {Aspuru-Guzik}},\ }\href {\doibase 10.1039/D1CS00932J}
  {\bibfield  {journal} {\bibinfo  {journal} {Chem. Soc. Rev.}\ }\textbf
  {\bibinfo {volume} {51}},\ \bibinfo {pages} {1659} (\bibinfo {year}
  {2022})},\ \Eprint {http://arxiv.org/abs/arxiv:2109.15176} {arxiv:2109.15176}
  \BibitemShut {NoStop}%
\bibitem [{\citenamefont {McArdle}\ \emph {et~al.}(2020)\citenamefont
  {McArdle}, \citenamefont {Endo}, \citenamefont {Aspuru-Guzik}, \citenamefont
  {Benjamin},\ and\ \citenamefont {Yuan}}]{McArdle2020}%
  \BibitemOpen
  \bibfield  {author} {\bibinfo {author} {\bibfnamefont {S.}~\bibnamefont
  {McArdle}}, \bibinfo {author} {\bibfnamefont {S.}~\bibnamefont {Endo}},
  \bibinfo {author} {\bibfnamefont {A.}~\bibnamefont {Aspuru-Guzik}}, \bibinfo
  {author} {\bibfnamefont {S.~C.}\ \bibnamefont {Benjamin}}, \ and\ \bibinfo
  {author} {\bibfnamefont {X.}~\bibnamefont {Yuan}},\ }\href {\doibase
  10.1103/RevModPhys.92.015003} {\bibfield  {journal} {\bibinfo  {journal}
  {Rev. Mod. Phys.}\ }\textbf {\bibinfo {volume} {92}},\ \bibinfo {pages}
  {015003} (\bibinfo {year} {2020})},\ \Eprint
  {http://arxiv.org/abs/arxiv:1808.10402} {arxiv:1808.10402} \BibitemShut
  {NoStop}%
\bibitem [{\citenamefont {Haidar}\ \emph {et~al.}(2022)\citenamefont {Haidar},
  \citenamefont {Ran{\v{c}}i{\'c}}, \citenamefont {Ayral}, \citenamefont
  {Maday},\ and\ \citenamefont {Piquemal}}]{haidar2022open}%
  \BibitemOpen
  \bibfield  {author} {\bibinfo {author} {\bibfnamefont {M.}~\bibnamefont
  {Haidar}}, \bibinfo {author} {\bibfnamefont {M.~J.}\ \bibnamefont
  {Ran{\v{c}}i{\'c}}}, \bibinfo {author} {\bibfnamefont {T.}~\bibnamefont
  {Ayral}}, \bibinfo {author} {\bibfnamefont {Y.}~\bibnamefont {Maday}}, \ and\
  \bibinfo {author} {\bibfnamefont {J.-P.}\ \bibnamefont {Piquemal}},\ }\href
  {\doibase 10.48550/ARXIV.2206.08798} {\enquote {\bibinfo {title} {{Open
  Source Variational Quantum Eigensolver Extension of the Quantum Learning
  Machine (QLM) for Quantum Chemistry}},}\ } (\bibinfo {year} {2022}),\
  \bibinfo {note} {https://arxiv.org/abs/2206.08798},\ \Eprint
  {http://arxiv.org/abs/arxiv:2206.08798} {arxiv:2206.08798} \BibitemShut
  {NoStop}%
\bibitem [{\citenamefont {Cade}\ \emph {et~al.}(2020)\citenamefont {Cade},
  \citenamefont {Mineh}, \citenamefont {Montanaro},\ and\ \citenamefont
  {Stanisic}}]{fh1}%
  \BibitemOpen
  \bibfield  {author} {\bibinfo {author} {\bibfnamefont {C.}~\bibnamefont
  {Cade}}, \bibinfo {author} {\bibfnamefont {L.}~\bibnamefont {Mineh}},
  \bibinfo {author} {\bibfnamefont {A.}~\bibnamefont {Montanaro}}, \ and\
  \bibinfo {author} {\bibfnamefont {S.}~\bibnamefont {Stanisic}},\ }\href
  {\doibase 10.1103/PhysRevB.102.235122} {\bibfield  {journal} {\bibinfo
  {journal} {Phys. Rev. B}\ }\textbf {\bibinfo {volume} {102}},\ \bibinfo
  {pages} {235122} (\bibinfo {year} {2020})},\ \Eprint
  {http://arxiv.org/abs/arxiv:1912.06007} {arxiv:1912.06007} \BibitemShut
  {NoStop}%
\bibitem [{\citenamefont {Cervera-Lierta}(2018)}]{cervera2018exact}%
  \BibitemOpen
  \bibfield  {author} {\bibinfo {author} {\bibfnamefont {A.}~\bibnamefont
  {Cervera-Lierta}},\ }\href {\doibase 10.22331/q-2018-12-21-114} {\bibfield
  {journal} {\bibinfo  {journal} {Quantum}\ }\textbf {\bibinfo {volume} {2}},\
  \bibinfo {pages} {114} (\bibinfo {year} {2018})},\ \Eprint
  {http://arxiv.org/abs/arxiv:1807.07112} {arxiv:1807.07112} \BibitemShut
  {NoStop}%
\bibitem [{\citenamefont {Cervia}\ \emph {et~al.}(2021)\citenamefont {Cervia},
  \citenamefont {Balantekin}, \citenamefont {Coppersmith}, \citenamefont
  {Johnson}, \citenamefont {Love}, \citenamefont {Poole}, \citenamefont
  {Robbins},\ and\ \citenamefont {Saffman}}]{cervia2021lipkin}%
  \BibitemOpen
  \bibfield  {author} {\bibinfo {author} {\bibfnamefont {M.~J.}\ \bibnamefont
  {Cervia}}, \bibinfo {author} {\bibfnamefont {A.}~\bibnamefont {Balantekin}},
  \bibinfo {author} {\bibfnamefont {S.}~\bibnamefont {Coppersmith}}, \bibinfo
  {author} {\bibfnamefont {C.~W.}\ \bibnamefont {Johnson}}, \bibinfo {author}
  {\bibfnamefont {P.~J.}\ \bibnamefont {Love}}, \bibinfo {author}
  {\bibfnamefont {C.}~\bibnamefont {Poole}}, \bibinfo {author} {\bibfnamefont
  {K.}~\bibnamefont {Robbins}}, \ and\ \bibinfo {author} {\bibfnamefont
  {M.}~\bibnamefont {Saffman}},\ }\href {\doibase 10.1103/PhysRevC.104.024305}
  {\bibfield  {journal} {\bibinfo  {journal} {Phys. Rev. C}\ }\textbf {\bibinfo
  {volume} {104}},\ \bibinfo {pages} {024305} (\bibinfo {year} {2021})},\
  \Eprint {http://arxiv.org/abs/arxiv:2011.04097} {arxiv:2011.04097}
  \BibitemShut {NoStop}%
\bibitem [{\citenamefont {Harsha}\ \emph {et~al.}(2018)\citenamefont {Harsha},
  \citenamefont {Shiozaki},\ and\ \citenamefont
  {Scuseria}}]{harsha2018difference}%
  \BibitemOpen
  \bibfield  {author} {\bibinfo {author} {\bibfnamefont {G.}~\bibnamefont
  {Harsha}}, \bibinfo {author} {\bibfnamefont {T.}~\bibnamefont {Shiozaki}}, \
  and\ \bibinfo {author} {\bibfnamefont {G.~E.}\ \bibnamefont {Scuseria}},\
  }\href {\doibase 10.1063/1.5011033} {\bibfield  {journal} {\bibinfo
  {journal} {J. Chem. Phys.}\ }\textbf {\bibinfo {volume} {148}},\ \bibinfo
  {pages} {044107} (\bibinfo {year} {2018})},\ \Eprint
  {http://arxiv.org/abs/arxiv:1711.00579} {arxiv:1711.00579} \BibitemShut
  {NoStop}%
\bibitem [{\citenamefont {Faba}\ \emph {et~al.}(2022)\citenamefont {Faba},
  \citenamefont {Mart\'{\i}n},\ and\ \citenamefont {Robledo}}]{faba3}%
  \BibitemOpen
  \bibfield  {author} {\bibinfo {author} {\bibfnamefont {J.}~\bibnamefont
  {Faba}}, \bibinfo {author} {\bibfnamefont {V.}~\bibnamefont {Mart\'{\i}n}}, \
  and\ \bibinfo {author} {\bibfnamefont {L.}~\bibnamefont {Robledo}},\ }\href
  {\doibase 10.1103/PhysRevA.105.062449} {\bibfield  {journal} {\bibinfo
  {journal} {Phys. Rev. A}\ }\textbf {\bibinfo {volume} {105}},\ \bibinfo
  {pages} {062449} (\bibinfo {year} {2022})},\ \Eprint
  {http://arxiv.org/abs/arxiv:2203.09400} {arxiv:2203.09400} \BibitemShut
  {NoStop}%
\bibitem [{\citenamefont {Wahlen-Strothman}\ \emph {et~al.}(2017)\citenamefont
  {Wahlen-Strothman}, \citenamefont {Henderson}, \citenamefont {Hermes},
  \citenamefont {Degroote}, \citenamefont {Qiu}, \citenamefont {Zhao},
  \citenamefont {Dukelsky},\ and\ \citenamefont
  {Scuseria}}]{wahlen2017merging}%
  \BibitemOpen
  \bibfield  {author} {\bibinfo {author} {\bibfnamefont {J.~M.}\ \bibnamefont
  {Wahlen-Strothman}}, \bibinfo {author} {\bibfnamefont {T.~M.}\ \bibnamefont
  {Henderson}}, \bibinfo {author} {\bibfnamefont {M.~R.}\ \bibnamefont
  {Hermes}}, \bibinfo {author} {\bibfnamefont {M.}~\bibnamefont {Degroote}},
  \bibinfo {author} {\bibfnamefont {Y.}~\bibnamefont {Qiu}}, \bibinfo {author}
  {\bibfnamefont {J.}~\bibnamefont {Zhao}}, \bibinfo {author} {\bibfnamefont
  {J.}~\bibnamefont {Dukelsky}}, \ and\ \bibinfo {author} {\bibfnamefont
  {G.~E.}\ \bibnamefont {Scuseria}},\ }\href {\doibase 10.1063/1.4974989}
  {\bibfield  {journal} {\bibinfo  {journal} {J. Chem. Phys.}\ }\textbf
  {\bibinfo {volume} {146}},\ \bibinfo {pages} {054110} (\bibinfo {year}
  {2017})},\ \Eprint {http://arxiv.org/abs/arxiv:1611.06273} {arxiv:1611.06273}
  \BibitemShut {NoStop}%
\bibitem [{\citenamefont {Robin}\ and\ \citenamefont
  {Savage}(2023)}]{Robin2023}%
  \BibitemOpen
  \bibfield  {author} {\bibinfo {author} {\bibfnamefont {C.~E.~P.}\
  \bibnamefont {Robin}}\ and\ \bibinfo {author} {\bibfnamefont {M.~J.}\
  \bibnamefont {Savage}},\ }\href {https://arxiv.org/abs/2301.05976} {\enquote
  {\bibinfo {title} {{Quantum Simulations in Effective Model Spaces (I):
  Hamiltonian Learning-VQE using Digital Quantum Computers and Application to
  the Lipkin-Meshkov-Glick Model}},}\ } (\bibinfo {year} {2023}),\ \bibinfo
  {note} {https://arxiv.org/abs/2301.05976},\ \Eprint
  {http://arxiv.org/abs/arxiv:2301.05976} {arxiv:2301.05976} \BibitemShut
  {NoStop}%
\bibitem [{\citenamefont {Lacroix}(2020)}]{ph1}%
  \BibitemOpen
  \bibfield  {author} {\bibinfo {author} {\bibfnamefont {D.}~\bibnamefont
  {Lacroix}},\ }\href {\doibase 10.1103/PhysRevLett.125.230502} {\bibfield
  {journal} {\bibinfo  {journal} {Phys. Rev. Lett.}\ }\textbf {\bibinfo
  {volume} {125}},\ \bibinfo {pages} {230502} (\bibinfo {year} {2020})},\
  \Eprint {http://arxiv.org/abs/arxiv:2006.06491} {arxiv:2006.06491}
  \BibitemShut {NoStop}%
\bibitem [{\citenamefont {Ruiz~Guzman}\ and\ \citenamefont
  {Lacroix}(2022)}]{ph2}%
  \BibitemOpen
  \bibfield  {author} {\bibinfo {author} {\bibfnamefont {E.~A.}\ \bibnamefont
  {Ruiz~Guzman}}\ and\ \bibinfo {author} {\bibfnamefont {D.}~\bibnamefont
  {Lacroix}},\ }\href {\doibase 10.1103/PhysRevC.105.024324} {\bibfield
  {journal} {\bibinfo  {journal} {Phys. Rev. C}\ }\textbf {\bibinfo {volume}
  {105}},\ \bibinfo {pages} {024324} (\bibinfo {year} {2022})},\ \Eprint
  {http://arxiv.org/abs/arxiv:2111.13080} {arxiv:2111.13080} \BibitemShut
  {NoStop}%
\bibitem [{\citenamefont {Qian}\ \emph {et~al.}(2022)\citenamefont {Qian},
  \citenamefont {Basili}, \citenamefont {Pal}, \citenamefont {Luecke},\ and\
  \citenamefont {Vary}}]{Qian:2021jxp}%
  \BibitemOpen
  \bibfield  {author} {\bibinfo {author} {\bibfnamefont {W.}~\bibnamefont
  {Qian}}, \bibinfo {author} {\bibfnamefont {R.}~\bibnamefont {Basili}},
  \bibinfo {author} {\bibfnamefont {S.}~\bibnamefont {Pal}}, \bibinfo {author}
  {\bibfnamefont {G.}~\bibnamefont {Luecke}}, \ and\ \bibinfo {author}
  {\bibfnamefont {J.~P.}\ \bibnamefont {Vary}},\ }\href {\doibase
  10.1103/PhysRevResearch.4.043193} {\bibfield  {journal} {\bibinfo  {journal}
  {Phys. Rev. Res.}\ }\textbf {\bibinfo {volume} {4}},\ \bibinfo {pages}
  {043193} (\bibinfo {year} {2022})},\ \Eprint
  {http://arxiv.org/abs/arxiv:2112.01927} {arXiv:arxiv:2112.01927 [quant-ph]}
  \BibitemShut {NoStop}%
\bibitem [{\citenamefont {Grimsley}\ \emph {et~al.}(2019)\citenamefont
  {Grimsley}, \citenamefont {Economou}, \citenamefont {Barnes},\ and\
  \citenamefont {Mayhall}}]{grimsley2019adaptive}%
  \BibitemOpen
  \bibfield  {author} {\bibinfo {author} {\bibfnamefont {H.~R.}\ \bibnamefont
  {Grimsley}}, \bibinfo {author} {\bibfnamefont {S.~E.}\ \bibnamefont
  {Economou}}, \bibinfo {author} {\bibfnamefont {E.}~\bibnamefont {Barnes}}, \
  and\ \bibinfo {author} {\bibfnamefont {N.~J.}\ \bibnamefont {Mayhall}},\
  }\href {\doibase 10.1038/s41467-019-10988-2} {\bibfield  {journal} {\bibinfo
  {journal} {Nat. Commun.}\ }\textbf {\bibinfo {volume} {10}},\ \bibinfo
  {pages} {1} (\bibinfo {year} {2019})},\ \Eprint
  {http://arxiv.org/abs/arxiv:1812.11173} {arxiv:1812.11173} \BibitemShut
  {NoStop}%
\bibitem [{\citenamefont {Sapova}\ and\ \citenamefont
  {Fedorov}(2022)}]{sapova22}%
  \BibitemOpen
  \bibfield  {author} {\bibinfo {author} {\bibfnamefont {M.~D.}\ \bibnamefont
  {Sapova}}\ and\ \bibinfo {author} {\bibfnamefont {A.~K.}\ \bibnamefont
  {Fedorov}},\ }\href {\doibase 10.1038/s42005-022-00982-4} {\bibfield
  {journal} {\bibinfo  {journal} {Commun. Phys.}\ }\textbf {\bibinfo {volume}
  {5}},\ \bibinfo {pages} {199} (\bibinfo {year} {2022})},\ \Eprint
  {http://arxiv.org/abs/arxiv:2108.11167} {arxiv:2108.11167} \BibitemShut
  {NoStop}%
\bibitem [{\citenamefont {Feniou}\ \emph {et~al.}(2023)\citenamefont {Feniou},
  \citenamefont {Hassan}, \citenamefont {Traoré}, \citenamefont {Giner},
  \citenamefont {Maday},\ and\ \citenamefont {Piquemal}}]{Feniou2023}%
  \BibitemOpen
  \bibfield  {author} {\bibinfo {author} {\bibfnamefont {C.}~\bibnamefont
  {Feniou}}, \bibinfo {author} {\bibfnamefont {M.}~\bibnamefont {Hassan}},
  \bibinfo {author} {\bibfnamefont {D.}~\bibnamefont {Traoré}}, \bibinfo
  {author} {\bibfnamefont {E.}~\bibnamefont {Giner}}, \bibinfo {author}
  {\bibfnamefont {Y.}~\bibnamefont {Maday}}, \ and\ \bibinfo {author}
  {\bibfnamefont {J.-P.}\ \bibnamefont {Piquemal}},\ }\href
  {https://arxiv.org/abs/2301.10196} {\enquote {\bibinfo {title}
  {{Overlap-ADAPT-VQE: Practical Quantum Chemistry on Quantum Computers via
  Overlap-Guided Compact Ans{\"a}tze}},}\ } (\bibinfo {year} {2023}),\ \bibinfo
  {note} {https://arxiv.org/abs/2301.10196},\ \Eprint
  {http://arxiv.org/abs/arxiv:2301.10196} {arxiv:2301.10196} \BibitemShut
  {NoStop}%
\bibitem [{\citenamefont {Dumitrescu}\ \emph {et~al.}(2018)\citenamefont
  {Dumitrescu}, \citenamefont {McCaskey}, \citenamefont {Hagen}, \citenamefont
  {Jansen}, \citenamefont {Morris}, \citenamefont {Papenbrock}, \citenamefont
  {Pooser}, \citenamefont {Dean},\ and\ \citenamefont
  {Lougovski}}]{Dumitrescu2018}%
  \BibitemOpen
  \bibfield  {author} {\bibinfo {author} {\bibfnamefont {E.~F.}\ \bibnamefont
  {Dumitrescu}}, \bibinfo {author} {\bibfnamefont {A.~J.}\ \bibnamefont
  {McCaskey}}, \bibinfo {author} {\bibfnamefont {G.}~\bibnamefont {Hagen}},
  \bibinfo {author} {\bibfnamefont {G.~R.}\ \bibnamefont {Jansen}}, \bibinfo
  {author} {\bibfnamefont {T.~D.}\ \bibnamefont {Morris}}, \bibinfo {author}
  {\bibfnamefont {T.}~\bibnamefont {Papenbrock}}, \bibinfo {author}
  {\bibfnamefont {R.~C.}\ \bibnamefont {Pooser}}, \bibinfo {author}
  {\bibfnamefont {D.~J.}\ \bibnamefont {Dean}}, \ and\ \bibinfo {author}
  {\bibfnamefont {P.}~\bibnamefont {Lougovski}},\ }\href {\doibase
  10.1103/PhysRevLett.120.210501} {\bibfield  {journal} {\bibinfo  {journal}
  {Phys. Rev. Lett.}\ }\textbf {\bibinfo {volume} {120}},\ \bibinfo {pages}
  {210501} (\bibinfo {year} {2018})},\ \Eprint
  {http://arxiv.org/abs/1801.03897} {1801.03897} \BibitemShut {NoStop}%
\bibitem [{\citenamefont {Lu}\ \emph {et~al.}(2019)\citenamefont {Lu},
  \citenamefont {Klco}, \citenamefont {Lukens}, \citenamefont {Morris},
  \citenamefont {Bansal}, \citenamefont {Ekstr\"om}, \citenamefont {Hagen},
  \citenamefont {Papenbrock}, \citenamefont {Weiner}, \citenamefont {Savage},\
  and\ \citenamefont {Lougovski}}]{LuKlco2019}%
  \BibitemOpen
  \bibfield  {author} {\bibinfo {author} {\bibfnamefont {H.-H.}\ \bibnamefont
  {Lu}}, \bibinfo {author} {\bibfnamefont {N.}~\bibnamefont {Klco}}, \bibinfo
  {author} {\bibfnamefont {J.~M.}\ \bibnamefont {Lukens}}, \bibinfo {author}
  {\bibfnamefont {T.~D.}\ \bibnamefont {Morris}}, \bibinfo {author}
  {\bibfnamefont {A.}~\bibnamefont {Bansal}}, \bibinfo {author} {\bibfnamefont
  {A.}~\bibnamefont {Ekstr\"om}}, \bibinfo {author} {\bibfnamefont
  {G.}~\bibnamefont {Hagen}}, \bibinfo {author} {\bibfnamefont
  {T.}~\bibnamefont {Papenbrock}}, \bibinfo {author} {\bibfnamefont {A.~M.}\
  \bibnamefont {Weiner}}, \bibinfo {author} {\bibfnamefont {M.~J.}\
  \bibnamefont {Savage}}, \ and\ \bibinfo {author} {\bibfnamefont
  {P.}~\bibnamefont {Lougovski}},\ }\href {\doibase
  10.1103/PhysRevA.100.012320} {\bibfield  {journal} {\bibinfo  {journal}
  {Phys. Rev. A}\ }\textbf {\bibinfo {volume} {100}},\ \bibinfo {pages}
  {012320} (\bibinfo {year} {2019})},\ \Eprint
  {http://arxiv.org/abs/arxiv:1810.03959} {arxiv:1810.03959} \BibitemShut
  {NoStop}%
\bibitem [{\citenamefont {Stetcu}\ \emph
  {et~al.}(2022{\natexlab{a}})\citenamefont {Stetcu}, \citenamefont {Baroni},\
  and\ \citenamefont {Carlson}}]{stetcu}%
  \BibitemOpen
  \bibfield  {author} {\bibinfo {author} {\bibfnamefont {I.}~\bibnamefont
  {Stetcu}}, \bibinfo {author} {\bibfnamefont {A.}~\bibnamefont {Baroni}}, \
  and\ \bibinfo {author} {\bibfnamefont {J.}~\bibnamefont {Carlson}},\ }\href
  {\doibase 10.1103/PhysRevC.105.064308} {\bibfield  {journal} {\bibinfo
  {journal} {Phys. Rev. C}\ }\textbf {\bibinfo {volume} {105}},\ \bibinfo
  {pages} {064308} (\bibinfo {year} {2022}{\natexlab{a}})},\ \Eprint
  {http://arxiv.org/abs/arxiv:2110.06098} {arxiv:2110.06098} \BibitemShut
  {NoStop}%
\bibitem [{\citenamefont {Kiss}\ \emph {et~al.}(2022)\citenamefont {Kiss},
  \citenamefont {Grossi}, \citenamefont {Lougovski}, \citenamefont {Sanchez},
  \citenamefont {Vallecorsa},\ and\ \citenamefont {Papenbrock}}]{papenbrock}%
  \BibitemOpen
  \bibfield  {author} {\bibinfo {author} {\bibfnamefont {O.}~\bibnamefont
  {Kiss}}, \bibinfo {author} {\bibfnamefont {M.}~\bibnamefont {Grossi}},
  \bibinfo {author} {\bibfnamefont {P.}~\bibnamefont {Lougovski}}, \bibinfo
  {author} {\bibfnamefont {F.}~\bibnamefont {Sanchez}}, \bibinfo {author}
  {\bibfnamefont {S.}~\bibnamefont {Vallecorsa}}, \ and\ \bibinfo {author}
  {\bibfnamefont {T.}~\bibnamefont {Papenbrock}},\ }\href {\doibase
  10.1103/PhysRevC.106.034325} {\bibfield  {journal} {\bibinfo  {journal}
  {Phys. Rev. C}\ }\textbf {\bibinfo {volume} {106}},\ \bibinfo {pages}
  {034325} (\bibinfo {year} {2022})},\ \Eprint
  {http://arxiv.org/abs/arxiv:2205.00864} {arxiv:2205.00864} \BibitemShut
  {NoStop}%
\bibitem [{\citenamefont {Shalit}\ and\ \citenamefont
  {Talmi}(1963)}]{Shalit1963}%
  \BibitemOpen
  \bibfield  {author} {\bibinfo {author} {\bibfnamefont {A.}~\bibnamefont
  {Shalit}}\ and\ \bibinfo {author} {\bibfnamefont {I.}~\bibnamefont {Talmi}},\
  }\href@noop {} {\emph {\bibinfo {title} {Nuclear Shell Theory}}}\ (\bibinfo
  {publisher} {Academic Press N. Y.},\ \bibinfo {year} {1963})\BibitemShut
  {NoStop}%
\bibitem [{\citenamefont {Talmi}(1993)}]{Talmi1993}%
  \BibitemOpen
  \bibfield  {author} {\bibinfo {author} {\bibfnamefont {I.}~\bibnamefont
  {Talmi}},\ }\href@noop {} {\emph {\bibinfo {title} {Simple Models of Complex
  Nuclei: The Shell Model and Interacting Boson Model}}},\ Beitrage Zur
  Wirtschaftsinformatik\ (\bibinfo  {publisher} {Harwood Academic Publishers},\
  \bibinfo {year} {1993})\BibitemShut {NoStop}%
\bibitem [{\citenamefont {Varshalovich}\ \emph {et~al.}(1988)\citenamefont
  {Varshalovich}, \citenamefont {Moskalev},\ and\ \citenamefont
  {Khersonskii}}]{varshalovich1988quantum}%
  \BibitemOpen
  \bibfield  {author} {\bibinfo {author} {\bibfnamefont {D.~A.}\ \bibnamefont
  {Varshalovich}}, \bibinfo {author} {\bibfnamefont {A.~N.}\ \bibnamefont
  {Moskalev}}, \ and\ \bibinfo {author} {\bibfnamefont {V.~K.}\ \bibnamefont
  {Khersonskii}},\ }\href@noop {} {\emph {\bibinfo {title} {Quantum theory of
  angular momentum}}}\ (\bibinfo  {publisher} {World Scientific},\ \bibinfo
  {year} {1988})\BibitemShut {NoStop}%
\bibitem [{\citenamefont {Hjorth-Jensen}\ \emph {et~al.}(1995)\citenamefont
  {Hjorth-Jensen}, \citenamefont {Kuo},\ and\ \citenamefont
  {Osnes}}]{HjorthJensen1995}%
  \BibitemOpen
  \bibfield  {author} {\bibinfo {author} {\bibfnamefont {M.}~\bibnamefont
  {Hjorth-Jensen}}, \bibinfo {author} {\bibfnamefont {T.~T.}\ \bibnamefont
  {Kuo}}, \ and\ \bibinfo {author} {\bibfnamefont {E.}~\bibnamefont {Osnes}},\
  }\href {\doibase 10.1016/0370-1573(95)00012-6} {\bibfield  {journal}
  {\bibinfo  {journal} {Phys. Rep.}\ }\textbf {\bibinfo {volume} {261}},\
  \bibinfo {pages} {125} (\bibinfo {year} {1995})}\BibitemShut {NoStop}%
\bibitem [{\citenamefont {Epelbaum}\ \emph {et~al.}(2009)\citenamefont
  {Epelbaum}, \citenamefont {Hammer},\ and\ \citenamefont
  {Meissner}}]{Epelbaum:2008ga}%
  \BibitemOpen
  \bibfield  {author} {\bibinfo {author} {\bibfnamefont {E.}~\bibnamefont
  {Epelbaum}}, \bibinfo {author} {\bibfnamefont {H.-W.}\ \bibnamefont
  {Hammer}}, \ and\ \bibinfo {author} {\bibfnamefont {U.-G.}\ \bibnamefont
  {Meissner}},\ }\href {\doibase 10.1103/RevModPhys.81.1773} {\bibfield
  {journal} {\bibinfo  {journal} {Rev. Mod. Phys.}\ }\textbf {\bibinfo {volume}
  {81}},\ \bibinfo {pages} {1773} (\bibinfo {year} {2009})},\ \Eprint
  {http://arxiv.org/abs/arxiv:0811.1338} {arXiv:arxiv:0811.1338 [nucl-th]}
  \BibitemShut {NoStop}%
\bibitem [{\citenamefont {Poves}\ and\ \citenamefont
  {Zuker}(1981)}]{Poves:1981zz}%
  \BibitemOpen
  \bibfield  {author} {\bibinfo {author} {\bibfnamefont {A.}~\bibnamefont
  {Poves}}\ and\ \bibinfo {author} {\bibfnamefont {A.}~\bibnamefont {Zuker}},\
  }\href {\doibase 10.1016/0370-1573(81)90153-8} {\bibfield  {journal}
  {\bibinfo  {journal} {Phys. Rep.}\ }\textbf {\bibinfo {volume} {70}},\
  \bibinfo {pages} {235} (\bibinfo {year} {1981})}\BibitemShut {NoStop}%
\bibitem [{\citenamefont {Cohen}\ and\ \citenamefont
  {Kurath}(1965)}]{cohen1965effective}%
  \BibitemOpen
  \bibfield  {author} {\bibinfo {author} {\bibfnamefont {S.}~\bibnamefont
  {Cohen}}\ and\ \bibinfo {author} {\bibfnamefont {D.}~\bibnamefont {Kurath}},\
  }\href {\doibase 10.1016/0029-5582(65)90148-3} {\bibfield  {journal}
  {\bibinfo  {journal} {Nuclear Physics}\ }\textbf {\bibinfo {volume} {73}},\
  \bibinfo {pages} {1} (\bibinfo {year} {1965})}\BibitemShut {NoStop}%
\bibitem [{\citenamefont {Brown}\ and\ \citenamefont
  {Richter}(2006)}]{Brown2006}%
  \BibitemOpen
  \bibfield  {author} {\bibinfo {author} {\bibfnamefont {B.~A.}\ \bibnamefont
  {Brown}}\ and\ \bibinfo {author} {\bibfnamefont {W.~A.}\ \bibnamefont
  {Richter}},\ }\href {\doibase 10.1103/PhysRevC.74.034315} {\bibfield
  {journal} {\bibinfo  {journal} {Phys. Rev. C}\ }\textbf {\bibinfo {volume}
  {74}},\ \bibinfo {pages} {034315} (\bibinfo {year} {2006})}\BibitemShut
  {NoStop}%
\bibitem [{\citenamefont {Poves}\ \emph {et~al.}(2001)\citenamefont {Poves},
  \citenamefont {Sánchez-Solano}, \citenamefont {Caurier},\ and\ \citenamefont
  {Nowacki}}]{Poves2001}%
  \BibitemOpen
  \bibfield  {author} {\bibinfo {author} {\bibfnamefont {A.}~\bibnamefont
  {Poves}}, \bibinfo {author} {\bibfnamefont {J.}~\bibnamefont
  {Sánchez-Solano}}, \bibinfo {author} {\bibfnamefont {E.}~\bibnamefont
  {Caurier}}, \ and\ \bibinfo {author} {\bibfnamefont {F.}~\bibnamefont
  {Nowacki}},\ }\href {\doibase 10.1016/S0375-9474(01)00967-8} {\bibfield
  {journal} {\bibinfo  {journal} {Nuclear Physics A}\ }\textbf {\bibinfo
  {volume} {694}},\ \bibinfo {pages} {157} (\bibinfo {year}
  {2001})}\BibitemShut {NoStop}%
\bibitem [{\citenamefont {Caurier}\ and\ \citenamefont
  {Nowacki}(1999)}]{caurier1999antoine}%
  \BibitemOpen
  \bibfield  {author} {\bibinfo {author} {\bibfnamefont {E.}~\bibnamefont
  {Caurier}}\ and\ \bibinfo {author} {\bibfnamefont {F.}~\bibnamefont
  {Nowacki}},\ }\href@noop {} {\bibfield  {journal} {\bibinfo  {journal} {Acta
  Physica Polonica}\ }\textbf {\bibinfo {volume} {30}},\ \bibinfo {pages} {705}
  (\bibinfo {year} {1999})}\BibitemShut {NoStop}%
\bibitem [{\citenamefont {Shimizu}\ \emph {et~al.}(2019)\citenamefont
  {Shimizu}, \citenamefont {Mizusaki}, \citenamefont {Utsuno},\ and\
  \citenamefont {Tsunoda}}]{Shimizu:2019xcd}%
  \BibitemOpen
  \bibfield  {author} {\bibinfo {author} {\bibfnamefont {N.}~\bibnamefont
  {Shimizu}}, \bibinfo {author} {\bibfnamefont {T.}~\bibnamefont {Mizusaki}},
  \bibinfo {author} {\bibfnamefont {Y.}~\bibnamefont {Utsuno}}, \ and\ \bibinfo
  {author} {\bibfnamefont {Y.}~\bibnamefont {Tsunoda}},\ }\href {\doibase
  10.1016/j.cpc.2019.06.011} {\bibfield  {journal} {\bibinfo  {journal}
  {Comput. Phys. Commun.}\ }\textbf {\bibinfo {volume} {244}},\ \bibinfo
  {pages} {372} (\bibinfo {year} {2019})},\ \Eprint
  {http://arxiv.org/abs/arxiv:1902.02064} {arXiv:arxiv:1902.02064 [nucl-th]}
  \BibitemShut {NoStop}%
\bibitem [{\citenamefont {Brown}\ and\ \citenamefont
  {Rae}(2014)}]{brown2014shell}%
  \BibitemOpen
  \bibfield  {author} {\bibinfo {author} {\bibfnamefont {B.}~\bibnamefont
  {Brown}}\ and\ \bibinfo {author} {\bibfnamefont {W.}~\bibnamefont {Rae}},\
  }\href {\doibase 10.101/j.nds.2014.07.022} {\bibfield  {journal} {\bibinfo
  {journal} {Nuclear Data Sheets}\ }\textbf {\bibinfo {volume} {120}},\
  \bibinfo {pages} {115} (\bibinfo {year} {2014})}\BibitemShut {NoStop}%
\bibitem [{\citenamefont {Johnson}\ \emph {et~al.}(2018)\citenamefont
  {Johnson}, \citenamefont {Ormand}, \citenamefont {McElvain},\ and\
  \citenamefont {Shan}}]{johnson2018bigstick}%
  \BibitemOpen
  \bibfield  {author} {\bibinfo {author} {\bibfnamefont {C.~W.}\ \bibnamefont
  {Johnson}}, \bibinfo {author} {\bibfnamefont {W.~E.}\ \bibnamefont {Ormand}},
  \bibinfo {author} {\bibfnamefont {K.~S.}\ \bibnamefont {McElvain}}, \ and\
  \bibinfo {author} {\bibfnamefont {H.}~\bibnamefont {Shan}},\ }\href
  {https://arxiv.org/abs/1801.08432} {\enquote {\bibinfo {title} {{BIGSTICK: A
  flexible configuration-interaction shell-model code}},}\ } (\bibinfo {year}
  {2018}),\ \bibinfo {note} {https://arxiv.org/abs/1801.08432},\ \Eprint
  {http://arxiv.org/abs/arxiv:1801.08432} {arxiv:1801.08432} \BibitemShut
  {NoStop}%
\bibitem [{\citenamefont {Seeley}\ \emph {et~al.}(2012)\citenamefont {Seeley},
  \citenamefont {Richard},\ and\ \citenamefont {Love}}]{Seeley2012}%
  \BibitemOpen
  \bibfield  {author} {\bibinfo {author} {\bibfnamefont {J.~T.}\ \bibnamefont
  {Seeley}}, \bibinfo {author} {\bibfnamefont {M.~J.}\ \bibnamefont {Richard}},
  \ and\ \bibinfo {author} {\bibfnamefont {P.~J.}\ \bibnamefont {Love}},\
  }\href {\doibase 10.1063/1.4768229} {\bibfield  {journal} {\bibinfo
  {journal} {J. Chem. Phys.}\ }\textbf {\bibinfo {volume} {137}},\ \bibinfo
  {pages} {224109} (\bibinfo {year} {2012})},\ \Eprint
  {http://arxiv.org/abs/arXiv:1208.5986} {arXiv:1208.5986} \BibitemShut
  {NoStop}%
\bibitem [{\citenamefont {Romero}\ \emph {et~al.}(2022)\citenamefont {Romero},
  \citenamefont {Engel}, \citenamefont {Tang},\ and\ \citenamefont
  {Economou}}]{romeroquantum}%
  \BibitemOpen
  \bibfield  {author} {\bibinfo {author} {\bibfnamefont {A.~M.}\ \bibnamefont
  {Romero}}, \bibinfo {author} {\bibfnamefont {J.}~\bibnamefont {Engel}},
  \bibinfo {author} {\bibfnamefont {H.~L.}\ \bibnamefont {Tang}}, \ and\
  \bibinfo {author} {\bibfnamefont {S.~E.}\ \bibnamefont {Economou}},\ }\href
  {\doibase 10.1103/PhysRevC.105.064317} {\bibfield  {journal} {\bibinfo
  {journal} {Phys. Rev. C}\ }\textbf {\bibinfo {volume} {105}},\ \bibinfo
  {pages} {064317} (\bibinfo {year} {2022})},\ \Eprint
  {http://arxiv.org/abs/arxiv:2203.01619} {arxiv:2203.01619} \BibitemShut
  {NoStop}%
\bibitem [{\citenamefont {Ritz}(1909)}]{ritz}%
  \BibitemOpen
  \bibfield  {author} {\bibinfo {author} {\bibfnamefont {W.}~\bibnamefont
  {Ritz}},\ }\href@noop {} {\bibfield  {journal} {\bibinfo  {journal} {Journal
  f{\"u}r die reine und angewandte Mathematik}\ }\textbf {\bibinfo {volume}
  {135}},\ \bibinfo {pages} {1} (\bibinfo {year} {1909})}\BibitemShut {NoStop}%
\bibitem [{\citenamefont {Rayleigh}(1870)}]{rayleigh}%
  \BibitemOpen
  \bibfield  {author} {\bibinfo {author} {\bibfnamefont {J.}~\bibnamefont
  {Rayleigh}},\ }\href@noop {} {\bibfield  {journal} {\bibinfo  {journal}
  {Phil. Trans}\ }\textbf {\bibinfo {volume} {161}},\ \bibinfo {pages} {16}
  (\bibinfo {year} {1870})}\BibitemShut {NoStop}%
\bibitem [{\citenamefont {Tang}\ \emph {et~al.}(2021)\citenamefont {Tang},
  \citenamefont {Shkolnikov}, \citenamefont {Barron}, \citenamefont {Grimsley},
  \citenamefont {Mayhall}, \citenamefont {Barnes},\ and\ \citenamefont
  {Economou}}]{tang2021qubit}%
  \BibitemOpen
  \bibfield  {author} {\bibinfo {author} {\bibfnamefont {H.~L.}\ \bibnamefont
  {Tang}}, \bibinfo {author} {\bibfnamefont {V.}~\bibnamefont {Shkolnikov}},
  \bibinfo {author} {\bibfnamefont {G.~S.}\ \bibnamefont {Barron}}, \bibinfo
  {author} {\bibfnamefont {H.~R.}\ \bibnamefont {Grimsley}}, \bibinfo {author}
  {\bibfnamefont {N.~J.}\ \bibnamefont {Mayhall}}, \bibinfo {author}
  {\bibfnamefont {E.}~\bibnamefont {Barnes}}, \ and\ \bibinfo {author}
  {\bibfnamefont {S.~E.}\ \bibnamefont {Economou}},\ }\href {\doibase
  10.1103/PRXQuantum.2.020310} {\bibfield  {journal} {\bibinfo  {journal} {PRX
  Quantum}\ }\textbf {\bibinfo {volume} {2}},\ \bibinfo {pages} {020310}
  (\bibinfo {year} {2021})},\ \Eprint {http://arxiv.org/abs/arxiv:1911.10205}
  {arxiv:1911.10205} \BibitemShut {NoStop}%
\bibitem [{\citenamefont {Barkoutsos}\ \emph {et~al.}(2018)\citenamefont
  {Barkoutsos}, \citenamefont {Gonthier}, \citenamefont {Sokolov},
  \citenamefont {Moll}, \citenamefont {Salis}, \citenamefont {Fuhrer},
  \citenamefont {Ganzhorn}, \citenamefont {Egger}, \citenamefont {Troyer},
  \citenamefont {Mezzacapo} \emph {et~al.}}]{barkoutsos2018quantum}%
  \BibitemOpen
  \bibfield  {author} {\bibinfo {author} {\bibfnamefont {P.~K.}\ \bibnamefont
  {Barkoutsos}}, \bibinfo {author} {\bibfnamefont {J.~F.}\ \bibnamefont
  {Gonthier}}, \bibinfo {author} {\bibfnamefont {I.}~\bibnamefont {Sokolov}},
  \bibinfo {author} {\bibfnamefont {N.}~\bibnamefont {Moll}}, \bibinfo {author}
  {\bibfnamefont {G.}~\bibnamefont {Salis}}, \bibinfo {author} {\bibfnamefont
  {A.}~\bibnamefont {Fuhrer}}, \bibinfo {author} {\bibfnamefont
  {M.}~\bibnamefont {Ganzhorn}}, \bibinfo {author} {\bibfnamefont {D.~J.}\
  \bibnamefont {Egger}}, \bibinfo {author} {\bibfnamefont {M.}~\bibnamefont
  {Troyer}}, \bibinfo {author} {\bibfnamefont {A.}~\bibnamefont {Mezzacapo}},
  \emph {et~al.},\ }\href {\doibase 10.1103/PhysRevA.98.022322} {\bibfield
  {journal} {\bibinfo  {journal} {Phys. Rev. A}\ }\textbf {\bibinfo {volume}
  {98}},\ \bibinfo {pages} {022322} (\bibinfo {year} {2018})},\ \Eprint
  {http://arxiv.org/abs/arxiv:1805.04340} {arxiv:1805.04340} \BibitemShut
  {NoStop}%
\bibitem [{\citenamefont {Childs}\ \emph {et~al.}(2021)\citenamefont {Childs},
  \citenamefont {Su}, \citenamefont {Tran}, \citenamefont {Wiebe},\ and\
  \citenamefont {Zhu}}]{childs2021theory}%
  \BibitemOpen
  \bibfield  {author} {\bibinfo {author} {\bibfnamefont {A.~M.}\ \bibnamefont
  {Childs}}, \bibinfo {author} {\bibfnamefont {Y.}~\bibnamefont {Su}}, \bibinfo
  {author} {\bibfnamefont {M.~C.}\ \bibnamefont {Tran}}, \bibinfo {author}
  {\bibfnamefont {N.}~\bibnamefont {Wiebe}}, \ and\ \bibinfo {author}
  {\bibfnamefont {S.}~\bibnamefont {Zhu}},\ }\href {\doibase
  10.1103/PhysRevX.11.011020} {\bibfield  {journal} {\bibinfo  {journal} {Phys.
  Rev. X}\ }\textbf {\bibinfo {volume} {11}},\ \bibinfo {pages} {011020}
  (\bibinfo {year} {2021})},\ \Eprint {http://arxiv.org/abs/arxiv:1912.08854}
  {arxiv:1912.08854} \BibitemShut {NoStop}%
\bibitem [{\citenamefont {Pellow-Jarman}\ \emph {et~al.}(2021)\citenamefont
  {Pellow-Jarman}, \citenamefont {Sinayskiy}, \citenamefont {Pillay},\ and\
  \citenamefont {Petruccione}}]{pellow2021comparison}%
  \BibitemOpen
  \bibfield  {author} {\bibinfo {author} {\bibfnamefont {A.}~\bibnamefont
  {Pellow-Jarman}}, \bibinfo {author} {\bibfnamefont {I.}~\bibnamefont
  {Sinayskiy}}, \bibinfo {author} {\bibfnamefont {A.}~\bibnamefont {Pillay}}, \
  and\ \bibinfo {author} {\bibfnamefont {F.}~\bibnamefont {Petruccione}},\
  }\href {\doibase 10.1007/s11128-021-03140-x} {\bibfield  {journal} {\bibinfo
  {journal} {Quantum Inf. Process.}\ }\textbf {\bibinfo {volume} {20}},\
  \bibinfo {pages} {1} (\bibinfo {year} {2021})},\ \Eprint
  {http://arxiv.org/abs/arxiv:2106.08682} {arxiv:2106.08682} \BibitemShut
  {NoStop}%
\bibitem [{\citenamefont {Stetcu}\ \emph
  {et~al.}(2022{\natexlab{b}})\citenamefont {Stetcu}, \citenamefont {Baroni},\
  and\ \citenamefont {Carlson}}]{Stetcu2023}%
  \BibitemOpen
  \bibfield  {author} {\bibinfo {author} {\bibfnamefont {I.}~\bibnamefont
  {Stetcu}}, \bibinfo {author} {\bibfnamefont {A.}~\bibnamefont {Baroni}}, \
  and\ \bibinfo {author} {\bibfnamefont {J.}~\bibnamefont {Carlson}},\ }\href
  {https://arxiv.org/abs/2211.10545} {\enquote {\bibinfo {title} {Projection
  algorithm for state preparation on quantum computers},}\ } (\bibinfo {year}
  {2022}{\natexlab{b}}),\ \bibinfo {note} {https://arxiv.org/abs/2211.10545},\
  \Eprint {http://arxiv.org/abs/arxiv:2211.10545} {arxiv:2211.10545}
  \BibitemShut {NoStop}%
\bibitem [{\citenamefont {Gigena}\ and\ \citenamefont
  {Rossignoli}(2015)}]{Gigena2015}%
  \BibitemOpen
  \bibfield  {author} {\bibinfo {author} {\bibfnamefont {N.}~\bibnamefont
  {Gigena}}\ and\ \bibinfo {author} {\bibfnamefont {R.}~\bibnamefont
  {Rossignoli}},\ }\href {\doibase 10.1103/PhysRevA.92.042326} {\bibfield
  {journal} {\bibinfo  {journal} {Phys. Rev. A}\ }\textbf {\bibinfo {volume}
  {92}},\ \bibinfo {pages} {042326} (\bibinfo {year} {2015})},\ \Eprint
  {http://arxiv.org/abs/arxiv:1509.05970} {arxiv:1509.05970} \BibitemShut
  {NoStop}%
\bibitem [{\citenamefont {Robin}\ \emph {et~al.}(2021)\citenamefont {Robin},
  \citenamefont {Savage},\ and\ \citenamefont {Pillet}}]{Robin2021}%
  \BibitemOpen
  \bibfield  {author} {\bibinfo {author} {\bibfnamefont {C.}~\bibnamefont
  {Robin}}, \bibinfo {author} {\bibfnamefont {M.~J.}\ \bibnamefont {Savage}}, \
  and\ \bibinfo {author} {\bibfnamefont {N.}~\bibnamefont {Pillet}},\ }\href
  {\doibase 10.1103/PhysRevC.103.034325} {\bibfield  {journal} {\bibinfo
  {journal} {Phys. Rev. C}\ }\textbf {\bibinfo {volume} {103}},\ \bibinfo
  {pages} {034325} (\bibinfo {year} {2021})},\ \Eprint
  {http://arxiv.org/abs/arxiv:2007.09157} {arxiv:2007.09157} \BibitemShut
  {NoStop}%
\bibitem [{\citenamefont {Johnson}\ and\ \citenamefont
  {Gorton}(2023)}]{johnson2023proton}%
  \BibitemOpen
  \bibfield  {author} {\bibinfo {author} {\bibfnamefont {C.~W.}\ \bibnamefont
  {Johnson}}\ and\ \bibinfo {author} {\bibfnamefont {O.~C.}\ \bibnamefont
  {Gorton}},\ }\href {\doibase 10.1088/1361-6471/acbece} {\bibfield  {journal}
  {\bibinfo  {journal} {Journal of Physics G: Nuclear and Particle Physics}\
  }\textbf {\bibinfo {volume} {50}},\ \bibinfo {pages} {045110} (\bibinfo
  {year} {2023})}\BibitemShut {NoStop}%
\bibitem [{\citenamefont {Bulgac}\ \emph {et~al.}(2023)\citenamefont {Bulgac},
  \citenamefont {Kafker},\ and\ \citenamefont
  {Abdurrahman}}]{bulgac2023measures}%
  \BibitemOpen
  \bibfield  {author} {\bibinfo {author} {\bibfnamefont {A.}~\bibnamefont
  {Bulgac}}, \bibinfo {author} {\bibfnamefont {M.}~\bibnamefont {Kafker}}, \
  and\ \bibinfo {author} {\bibfnamefont {I.}~\bibnamefont {Abdurrahman}},\
  }\href {\doibase 10.1103/PhysRevC.107.044318} {\bibfield  {journal} {\bibinfo
   {journal} {Phys. Rev. C}\ }\textbf {\bibinfo {volume} {107}},\ \bibinfo
  {pages} {044318} (\bibinfo {year} {2023})}\BibitemShut {NoStop}%
\bibitem [{\citenamefont {Pazy}(2023)}]{Pazy2023}%
  \BibitemOpen
  \bibfield  {author} {\bibinfo {author} {\bibfnamefont {E.}~\bibnamefont
  {Pazy}},\ }\href {\doibase 10.1103/PhysRevC.107.054308} {\bibfield  {journal}
  {\bibinfo  {journal} {Phys. Rev. C}\ }\textbf {\bibinfo {volume} {107}},\
  \bibinfo {pages} {054308} (\bibinfo {year} {2023})}\BibitemShut {NoStop}%
\bibitem [{\citenamefont {Bulgac}(2022)}]{Bulgac2022b}%
  \BibitemOpen
  \bibfield  {author} {\bibinfo {author} {\bibfnamefont {A.}~\bibnamefont
  {Bulgac}},\ }\href {https://arxiv.org/abs/2203.12079} {\enquote {\bibinfo
  {title} {Entanglement entropy, single-particle occupation probabilities, and
  short-range correlations},}\ } (\bibinfo {year} {2022}),\ \bibinfo {note}
  {https://arxiv.org/abs/2203.12079},\ \Eprint
  {http://arxiv.org/abs/arxiv:2203.12079} {arxiv:2203.12079} \BibitemShut
  {NoStop}%
\bibitem [{\citenamefont {Lee}\ \emph {et~al.}(2023)\citenamefont {Lee},
  \citenamefont {Lee}, \citenamefont {Zhai}, \citenamefont {Tong},
  \citenamefont {Dalzell}, \citenamefont {Kumar}, \citenamefont {Helms},
  \citenamefont {Gray}, \citenamefont {Cui}, \citenamefont {Liu} \emph
  {et~al.}}]{lee2023evaluating}%
  \BibitemOpen
  \bibfield  {author} {\bibinfo {author} {\bibfnamefont {S.}~\bibnamefont
  {Lee}}, \bibinfo {author} {\bibfnamefont {J.}~\bibnamefont {Lee}}, \bibinfo
  {author} {\bibfnamefont {H.}~\bibnamefont {Zhai}}, \bibinfo {author}
  {\bibfnamefont {Y.}~\bibnamefont {Tong}}, \bibinfo {author} {\bibfnamefont
  {A.~M.}\ \bibnamefont {Dalzell}}, \bibinfo {author} {\bibfnamefont
  {A.}~\bibnamefont {Kumar}}, \bibinfo {author} {\bibfnamefont
  {P.}~\bibnamefont {Helms}}, \bibinfo {author} {\bibfnamefont
  {J.}~\bibnamefont {Gray}}, \bibinfo {author} {\bibfnamefont {Z.-H.}\
  \bibnamefont {Cui}}, \bibinfo {author} {\bibfnamefont {W.}~\bibnamefont
  {Liu}},  \emph {et~al.},\ }\href {\doibase 10.1038/s41467-023-37587-6}
  {\bibfield  {journal} {\bibinfo  {journal} {Nature Communications}\ }\textbf
  {\bibinfo {volume} {14}},\ \bibinfo {pages} {1952} (\bibinfo {year}
  {2023})}\BibitemShut {NoStop}%
\bibitem [{\citenamefont {Di~Matteo}\ \emph {et~al.}(2021)\citenamefont
  {Di~Matteo}, \citenamefont {McCoy}, \citenamefont {Gysbers}, \citenamefont
  {Miyagi}, \citenamefont {Woloshyn},\ and\ \citenamefont
  {Navr{\'a}til}}]{di2021improving}%
  \BibitemOpen
  \bibfield  {author} {\bibinfo {author} {\bibfnamefont {O.}~\bibnamefont
  {Di~Matteo}}, \bibinfo {author} {\bibfnamefont {A.}~\bibnamefont {McCoy}},
  \bibinfo {author} {\bibfnamefont {P.}~\bibnamefont {Gysbers}}, \bibinfo
  {author} {\bibfnamefont {T.}~\bibnamefont {Miyagi}}, \bibinfo {author}
  {\bibfnamefont {R.}~\bibnamefont {Woloshyn}}, \ and\ \bibinfo {author}
  {\bibfnamefont {P.}~\bibnamefont {Navr{\'a}til}},\ }\href {\doibase
  10.1103/PhysRevA.103.042405} {\bibfield  {journal} {\bibinfo  {journal}
  {Phys. Rev. A}\ }\textbf {\bibinfo {volume} {103}},\ \bibinfo {pages}
  {042405} (\bibinfo {year} {2021})},\ \Eprint
  {http://arxiv.org/abs/arxiv:2008.05012} {arxiv:2008.05012} \BibitemShut
  {NoStop}%
\bibitem [{\citenamefont {Siwach}\ and\ \citenamefont
  {Arumugam}(2021)}]{siwach2021quantum}%
  \BibitemOpen
  \bibfield  {author} {\bibinfo {author} {\bibfnamefont {P.}~\bibnamefont
  {Siwach}}\ and\ \bibinfo {author} {\bibfnamefont {P.}~\bibnamefont
  {Arumugam}},\ }\href {\doibase 10.1103/PhysRevC.104.034301} {\bibfield
  {journal} {\bibinfo  {journal} {Phys. Rev. C}\ }\textbf {\bibinfo {volume}
  {104}},\ \bibinfo {pages} {034301} (\bibinfo {year} {2021})}\BibitemShut
  {NoStop}%
\bibitem [{\citenamefont {Faba}\ \emph
  {et~al.}(2021{\natexlab{a}})\citenamefont {Faba}, \citenamefont
  {Mart{\'\i}n},\ and\ \citenamefont {Robledo}}]{faba2021correlation}%
  \BibitemOpen
  \bibfield  {author} {\bibinfo {author} {\bibfnamefont {J.}~\bibnamefont
  {Faba}}, \bibinfo {author} {\bibfnamefont {V.}~\bibnamefont {Mart{\'\i}n}}, \
  and\ \bibinfo {author} {\bibfnamefont {L.}~\bibnamefont {Robledo}},\ }\href
  {\doibase 10.1103/PhysRevA.104.032428} {\bibfield  {journal} {\bibinfo
  {journal} {Phys. Rev. A}\ }\textbf {\bibinfo {volume} {104}},\ \bibinfo
  {pages} {032428} (\bibinfo {year} {2021}{\natexlab{a}})},\ \Eprint
  {http://arxiv.org/abs/arxiv:2106.15993} {arxiv:2106.15993} \BibitemShut
  {NoStop}%
\bibitem [{\citenamefont {Faba}\ \emph
  {et~al.}(2021{\natexlab{b}})\citenamefont {Faba}, \citenamefont
  {Mart{\'\i}n},\ and\ \citenamefont {Robledo}}]{faba2021two}%
  \BibitemOpen
  \bibfield  {author} {\bibinfo {author} {\bibfnamefont {J.}~\bibnamefont
  {Faba}}, \bibinfo {author} {\bibfnamefont {V.}~\bibnamefont {Mart{\'\i}n}}, \
  and\ \bibinfo {author} {\bibfnamefont {L.}~\bibnamefont {Robledo}},\ }\href
  {\doibase 10.1103/PhysRevA.103.032426} {\bibfield  {journal} {\bibinfo
  {journal} {Phys. Rev. A}\ }\textbf {\bibinfo {volume} {103}},\ \bibinfo
  {pages} {032426} (\bibinfo {year} {2021}{\natexlab{b}})},\ \Eprint
  {http://arxiv.org/abs/arxiv:2012.01292} {arxiv:2012.01292} \BibitemShut
  {NoStop}%
\bibitem [{\citenamefont {Kirby}\ \emph {et~al.}(2023)\citenamefont {Kirby},
  \citenamefont {Motta},\ and\ \citenamefont {Mezzacapo}}]{kirby2023exact}%
  \BibitemOpen
  \bibfield  {author} {\bibinfo {author} {\bibfnamefont {W.}~\bibnamefont
  {Kirby}}, \bibinfo {author} {\bibfnamefont {M.}~\bibnamefont {Motta}}, \ and\
  \bibinfo {author} {\bibfnamefont {A.}~\bibnamefont {Mezzacapo}},\ }\href
  {\doibase 10.22331/q-2023-05-23-1018} {\bibfield  {journal} {\bibinfo
  {journal} {Quantum}\ }\textbf {\bibinfo {volume} {7}},\ \bibinfo {pages}
  {1018} (\bibinfo {year} {2023})}\BibitemShut {NoStop}%
\bibitem [{\citenamefont {Efthymiou}\ \emph {et~al.}(2021)\citenamefont
  {Efthymiou}, \citenamefont {Ramos-Calderer}, \citenamefont {Bravo-Prieto},
  \citenamefont {P\'erez-Salinas}, \citenamefont {Garc\'ia-Mart\'in},
  \citenamefont {Garcia-Saez}, \citenamefont {Latorre},\ and\ \citenamefont
  {Carrazza}}]{Efthymiou_2022}%
  \BibitemOpen
  \bibfield  {author} {\bibinfo {author} {\bibfnamefont {S.}~\bibnamefont
  {Efthymiou}}, \bibinfo {author} {\bibfnamefont {S.}~\bibnamefont
  {Ramos-Calderer}}, \bibinfo {author} {\bibfnamefont {C.}~\bibnamefont
  {Bravo-Prieto}}, \bibinfo {author} {\bibfnamefont {A.}~\bibnamefont
  {P\'erez-Salinas}}, \bibinfo {author} {\bibfnamefont {D.}~\bibnamefont
  {Garc\'ia-Mart\'in}}, \bibinfo {author} {\bibfnamefont {A.}~\bibnamefont
  {Garcia-Saez}}, \bibinfo {author} {\bibfnamefont {J.~I.}\ \bibnamefont
  {Latorre}}, \ and\ \bibinfo {author} {\bibfnamefont {S.}~\bibnamefont
  {Carrazza}},\ }\href {\doibase 10.1088/2058-9565/ac39f5} {\bibfield
  {journal} {\bibinfo  {journal} {Quantum Sci. Technol.}\ }\textbf {\bibinfo
  {volume} {7}},\ \bibinfo {pages} {015018} (\bibinfo {year} {2021})},\ \Eprint
  {http://arxiv.org/abs/arxiv:2009.01845} {arxiv:2009.01845} \BibitemShut
  {NoStop}%
\bibitem [{\citenamefont {Lam}\ \emph {et~al.}(2015)\citenamefont {Lam},
  \citenamefont {Pitrou},\ and\ \citenamefont {Seibert}}]{lam15}%
  \BibitemOpen
  \bibfield  {author} {\bibinfo {author} {\bibfnamefont {S.~K.}\ \bibnamefont
  {Lam}}, \bibinfo {author} {\bibfnamefont {A.}~\bibnamefont {Pitrou}}, \ and\
  \bibinfo {author} {\bibfnamefont {S.}~\bibnamefont {Seibert}},\ }in\ \href
  {\doibase 10.1145/2833157.2833162} {\emph {\bibinfo {booktitle} {Proceedings
  of the Second Workshop on the LLVM Compiler Infrastructure in HPC}}},\
  \bibinfo {series and number} {LLVM '15}\ (\bibinfo  {publisher} {Association
  for Computing Machinery},\ \bibinfo {address} {New York, NY, USA},\ \bibinfo
  {year} {2015})\BibitemShut {NoStop}%
\bibitem [{\citenamefont {P\'erez-Obiol}\ \emph {et~al.}(2022)\citenamefont
  {P\'erez-Obiol}, \citenamefont {P\'erez-Salinas}, \citenamefont
  {S\'anchez-Ram\'{\i}rez}, \citenamefont {Ara\'ujo},\ and\ \citenamefont
  {Garcia-Saez}}]{perezobiol_2022}%
  \BibitemOpen
  \bibfield  {author} {\bibinfo {author} {\bibfnamefont {A.}~\bibnamefont
  {P\'erez-Obiol}}, \bibinfo {author} {\bibfnamefont {A.}~\bibnamefont
  {P\'erez-Salinas}}, \bibinfo {author} {\bibfnamefont {S.}~\bibnamefont
  {S\'anchez-Ram\'{\i}rez}}, \bibinfo {author} {\bibfnamefont {B.~G.~M.}\
  \bibnamefont {Ara\'ujo}}, \ and\ \bibinfo {author} {\bibfnamefont
  {A.}~\bibnamefont {Garcia-Saez}},\ }\href {\doibase
  10.1103/PhysRevA.106.052408} {\bibfield  {journal} {\bibinfo  {journal}
  {Phys. Rev. A}\ }\textbf {\bibinfo {volume} {106}},\ \bibinfo {pages}
  {052408} (\bibinfo {year} {2022})},\ \Eprint
  {http://arxiv.org/abs/arxiv:2204.03013} {arxiv:2204.03013} \BibitemShut
  {NoStop}%
\bibitem [{\citenamefont {Okuta}\ \emph {et~al.}(2017)\citenamefont {Okuta},
  \citenamefont {Unno}, \citenamefont {Nishino}, \citenamefont {Hido},\ and\
  \citenamefont {Loomis}}]{okuta17cupy}%
  \BibitemOpen
  \bibfield  {author} {\bibinfo {author} {\bibfnamefont {R.}~\bibnamefont
  {Okuta}}, \bibinfo {author} {\bibfnamefont {Y.}~\bibnamefont {Unno}},
  \bibinfo {author} {\bibfnamefont {D.}~\bibnamefont {Nishino}}, \bibinfo
  {author} {\bibfnamefont {S.}~\bibnamefont {Hido}}, \ and\ \bibinfo {author}
  {\bibfnamefont {C.}~\bibnamefont {Loomis}},\ }in\ \href
  {http://learningsys.org/nips17/assets/papers/paper_16.pdf} {\emph {\bibinfo
  {booktitle} {Proceedings of Workshop on ML Systems in The Thirty-first Annual
  Conference on Neural Information Processing Systems (NIPS)}}}\ (\bibinfo
  {year} {2017})\BibitemShut {NoStop}%
\bibitem [{\citenamefont {Jordan}\ and\ \citenamefont {Wigner}(1993)}]{jw}%
  \BibitemOpen
  \bibfield  {author} {\bibinfo {author} {\bibfnamefont {P.}~\bibnamefont
  {Jordan}}\ and\ \bibinfo {author} {\bibfnamefont {E.~P.}\ \bibnamefont
  {Wigner}},\ }in\ \href@noop {} {\emph {\bibinfo {booktitle} {The Collected
  Works of Eugene Paul Wigner}}}\ (\bibinfo  {publisher} {Springer},\ \bibinfo
  {year} {1993})\ pp.\ \bibinfo {pages} {109--129}\BibitemShut {NoStop}%
\bibitem [{\citenamefont {McClean}\ \emph {et~al.}(2020)\citenamefont
  {McClean}, \citenamefont {Rubin}, \citenamefont {Sung}, \citenamefont
  {Kivlichan}, \citenamefont {Bonet-Monroig}, \citenamefont {Cao},
  \citenamefont {Dai}, \citenamefont {Fried}, \citenamefont {Gidney},
  \citenamefont {Gimby} \emph {et~al.}}]{openfermion}%
  \BibitemOpen
  \bibfield  {author} {\bibinfo {author} {\bibfnamefont {J.~R.}\ \bibnamefont
  {McClean}}, \bibinfo {author} {\bibfnamefont {N.~C.}\ \bibnamefont {Rubin}},
  \bibinfo {author} {\bibfnamefont {K.~J.}\ \bibnamefont {Sung}}, \bibinfo
  {author} {\bibfnamefont {I.~D.}\ \bibnamefont {Kivlichan}}, \bibinfo {author}
  {\bibfnamefont {X.}~\bibnamefont {Bonet-Monroig}}, \bibinfo {author}
  {\bibfnamefont {Y.}~\bibnamefont {Cao}}, \bibinfo {author} {\bibfnamefont
  {C.}~\bibnamefont {Dai}}, \bibinfo {author} {\bibfnamefont {E.~S.}\
  \bibnamefont {Fried}}, \bibinfo {author} {\bibfnamefont {C.}~\bibnamefont
  {Gidney}}, \bibinfo {author} {\bibfnamefont {B.}~\bibnamefont {Gimby}},
  \emph {et~al.},\ }\href {\doibase 10.1088/2058-9565/ab8ebc} {\bibfield
  {journal} {\bibinfo  {journal} {Quantum Sci. Technol.}\ }\textbf {\bibinfo
  {volume} {5}},\ \bibinfo {pages} {034014} (\bibinfo {year} {2020})},\ \Eprint
  {http://arxiv.org/abs/arxiv:1710.07629} {arxiv:1710.07629} \BibitemShut
  {NoStop}%
\bibitem [{\citenamefont {Sawaya}\ \emph {et~al.}(2020)\citenamefont {Sawaya},
  \citenamefont {Menke}, \citenamefont {Kyaw}, \citenamefont {Johri},
  \citenamefont {Aspuru-Guzik},\ and\ \citenamefont {Guerreschi}}]{sawaya20}%
  \BibitemOpen
  \bibfield  {author} {\bibinfo {author} {\bibfnamefont {N.~P.~D.}\
  \bibnamefont {Sawaya}}, \bibinfo {author} {\bibfnamefont {T.}~\bibnamefont
  {Menke}}, \bibinfo {author} {\bibfnamefont {T.~H.}\ \bibnamefont {Kyaw}},
  \bibinfo {author} {\bibfnamefont {S.}~\bibnamefont {Johri}}, \bibinfo
  {author} {\bibfnamefont {A.}~\bibnamefont {Aspuru-Guzik}}, \ and\ \bibinfo
  {author} {\bibfnamefont {G.~G.}\ \bibnamefont {Guerreschi}},\ }\href
  {\doibase 10.1038/s41534-020-0278-0} {\bibfield  {journal} {\bibinfo
  {journal} {npj Quantum Inf.}\ }\textbf {\bibinfo {volume} {6}},\ \bibinfo
  {pages} {49} (\bibinfo {year} {2020})},\ \Eprint
  {http://arxiv.org/abs/arxiv:1909.12847} {arxiv:1909.12847} \BibitemShut
  {NoStop}%
\bibitem [{\citenamefont {Carbone}\ \emph {et~al.}(2013)\citenamefont
  {Carbone}, \citenamefont {Cipollone}, \citenamefont {Barbieri}, \citenamefont
  {Rios},\ and\ \citenamefont {Polls}}]{Carbone2013}%
  \BibitemOpen
  \bibfield  {author} {\bibinfo {author} {\bibfnamefont {A.}~\bibnamefont
  {Carbone}}, \bibinfo {author} {\bibfnamefont {A.}~\bibnamefont {Cipollone}},
  \bibinfo {author} {\bibfnamefont {C.}~\bibnamefont {Barbieri}}, \bibinfo
  {author} {\bibfnamefont {A.}~\bibnamefont {Rios}}, \ and\ \bibinfo {author}
  {\bibfnamefont {A.}~\bibnamefont {Polls}},\ }\href {\doibase
  10.1103/PhysRevC.88.054326} {\bibfield  {journal} {\bibinfo  {journal} {Phys.
  Rev. C}\ }\textbf {\bibinfo {volume} {88}},\ \bibinfo {pages} {054326}
  (\bibinfo {year} {2013})}\BibitemShut {NoStop}%
\bibitem [{\citenamefont {Hebeler}(2021)}]{Hebeler:2020ocj}%
  \BibitemOpen
  \bibfield  {author} {\bibinfo {author} {\bibfnamefont {K.}~\bibnamefont
  {Hebeler}},\ }\href {\doibase 10.1016/j.physrep.2020.08.009} {\bibfield
  {journal} {\bibinfo  {journal} {Physics Reports}\ }\textbf {\bibinfo {volume}
  {890}},\ \bibinfo {pages} {1} (\bibinfo {year} {2021})},\ \Eprint
  {http://arxiv.org/abs/arxiv:2002.09548} {arXiv:arxiv:2002.09548 [nucl-th]}
  \BibitemShut {NoStop}%
\end{thebibliography}%


\section*{Acknowledgements}
A.M.R. thanks J. Engel for the support and fruitful discussions in the conception of this project. A. P-O. and A. G-S. thank the QUANTIC group at BSC for insightful comments and discussions along the realization of this work. A. G-S. received funding from the European Union’s Horizon 2020 research and innovation programme under grant agreement No 951911 (AI4Media). This work is financially supported by 
the Ministry of Economic Affairs and Digital Transformation of the Spanish Government through the QUANTUM ENIA project call - Quantum Spain project, 
by the European Union through the Recovery, Transformation and Resilience Plan - NextGenerationEU within the framework of the Digital Spain 2026 Agenda,
by grants PID2020-118758GB-I00 and PID2020-114626GB-I00 
funded by MCIN/AEI/10.13039/501100011033; 
by the "Ram\'on y Cajal" grants RYC-2017-22781 and RYC2018-026072 funded by MCIN/AEI /10.13039/501100011033 and FSE “El FSE invierte en tu futuro”; and 
by the  “Unit of Excellence Mar\'ia de Maeztu 2020-2023” award to the Institute of Cosmos Sciences, Grant CEX2019-000918-M funded by MCIN/AEI/10.13039/501100011033.

\section*{Author contributions statement}
A.P.O. and A.M.R. designed and performed research; A.P.O. analyzed data; and A.P.O., A.M.R., J.M., A.R., A.G.S., and B.J.D. wrote the paper.


\section*{Additional information} 

{\bf Supplementary information} 
The online version contains supplementary material.

\noindent {\bf Competing interests} 
The authors declare no competing interests.

\noindent {\bf Correspondence} and requests for materials should be addressed to AMR or APO.

\section*{Supplementary information}

\section*{Circuit design strategy}

\subsection*{Number of different measurement circuits}

Here we discuss the number of different 
measurement circuits that are necessary 
to compute
expectation values of the energy as well
as of the products of operators required
in the gradient calculations. 
Local terms $n_i$ and $h_{ijij}$ can be measured simultaneously. 
We analyze and optimize the number of different circuits needed to measure the expectation value of the non-local part of $H_{\rm{eff}}$, $h_{ijki}$ and $h_{ijkl}$, for the $p$, $sd$ and $pf$ shell valence spaces.

All terms $h_{ijki}=-n_i(a_j^\dagger a_k +a_k^\dagger a_j)$ with the same hopping (same indices j, k) and different local terms $n_i$ can be measured simultaneously since they commute, $[h_{ijik}, h_{i'ji'k}]=0$. The local part of $h_{ijik}$ conserves the third components of the angular momentum and isospin, $m$ and $t_z$, implying that the complementary hopping term involves only indices in the same vertical axis in the  panel (b) diagram of Fig.~1 in the main text. 
For example, considering the $sd$ shell with only neutrons, this amounts to a total of eight terms:
\begin{equation}
\{(j,k)\} = \{(1,8), (2,6), (6,9), (2,9),(3,7), (7,10), (3,10), (4,11)\}.        
\end{equation}
The number of different circuits needed to measure all $h_{ijki}$ terms is then equivalent to the number of different $m$- and $t_z$-conserving single-excitation operators in the shell. This scales, in the worst case, as $O(N_{qb}^2)$, representing a relatively small number of circuits. Each term is diagonalized with the circuit $M_{jk}=CX_{kj}H_k CX_{kj}$, which for continguous indices, $k=j+1$, results in the operator $|101\rangle\langle 101|-|110\rangle\langle 110|$, where the indices $(i,j,k)$ have been omitted.
Therefore, $\langle h_{ijik}\rangle = 
p_{101}^{(ijk)}-p_{110}^{(ijk)}$, with $p_{101}^{(ijk)}$ and $p_{110}^{(ijk)}$ the probabilities of measuring $101$ and $110$ in qubits $(i,j,k)$ after the change of basis.

The double-hopping terms $h_{ijkl}$ that involve different sets of orbitals ($i$, $j$, $k$, $l$) also commute and can be measured with the same circuit. Given a group of self-commuting terms, products of $Z$s of one or more terms $h_{ijkl}$ appearing in the JW mapping may overlap with the indices of another term $h_{i'j'k'l'}$ in the group. A product of an even number of overlapping $Zs$, for example $P_{\textrm{even}}=Z_{i'}Z_{j'}$, commutes with $M_{i'j'k'l'}$ and the same circuit $M_{ijkl}$ can be used for both. If there is a product of an odd number of overlapping $Z$s, $P_{\textrm{odd}}$, then $[P_{odd},M_{i'j'k'l'}]\neq0$ and all the different $h_{ijkl}$ operators need to be diagonalized simultaneously. 
Some terms that share two indices also commute, but for simplicity we do not group them into the same measurement. 

\subsection{Simultaneous diagonalization of double-hopping terms with different indices}
\label{app:diagonalization}

Measuring the expected value of the Hamiltonian requires then a simultaneous diagonalization of each term $h_{ijkl}$ with different values for the indices ($i$, $j$, $k$, $l$). These operators consist of the product $h_{ijkl}=P_{ij}^{kl}O_{ijkl}$, where $P_{ij}^{kl}$ is a diagonal Pauli string containing only $Z$s and $O_{ijkl}$ is the non-diagonal part,
\begin{equation}
\begin{split}
O_{ijkl} \equiv\,&
\left(\sigma^-_i\sigma^-_j\sigma^+_k\sigma^+_l + \sigma^-_k\sigma^-_l\sigma^+_i\sigma^+_j\right)
\\=&
\frac18(
X_i X_j X_k X_l - X_i X_j Y_k Y_l + X_i Y_j X_k Y_l \\&+ X_i Y_j Y_k X_l + 
Y_i Y_j Y_k Y_l - Y_i Y_j X_k X_l \\&+ Y_i X_j Y_k X_l + Y_i X_j X_k Y_l 
),
\\=&
|0011\rangle\langle 1100|+|1100\rangle\langle 0011|,
\end{split}
\end{equation}
where in the last line the indices ($i,j,k,l$) have been omitted, see Table~1 in the main text. To diagonalize a single term $h_{ijkl}$ we use the change of basis $M_{ijkl}\equiv\, CX_{ij}CX_{ki}CX_{lk} H_l CX_{lk} CX_{ki} CX_{ij}$, such that
\begin{equation}
\begin{split}
M_{ijkl}^\dagger O_{ijkl} M_{ijkl}\equiv\,&
D_{ijkl}=
|1100\rangle\langle 1100|-|0011\rangle\langle 0011|.
\end{split}
\end{equation}

For contiguous indices, $j=i+1$, $l=k+1$, then $P_{ij}^{kl}=1$,
and we have $\langle h_{ijkl}\rangle = p_{1100}^{(ijkl)}-p_{0011}^{(ijkl)}$, dependent on the probabilities of measuring 1100 and 0011 in qubits ($i$, $j$, $k$, $l$) after applying the change of basis, as stated in Eq.~(18) in the main text.
In the general case, $j>i+1$, $l>k+1$, and $P_{ij}^{kl}\neq1$,
the expected value needs to account for the product of $Z$ matrices.
For example, considering $\langle Z_q\rangle=p_0^{(q)}-p_1^{(q)}$ and
 $\langle Z_qZ_r\rangle=p_{00}^{(qr)}-p_{01}^{(qr)}-p_{10}^{(qr)}+p_{11}^{(qr)}$,
 \begin{equation}
 \begin{split}
 \langle Z_q O_{ijkl}\rangle =&
 [p_{01100}^{(qijkl)}-p_{00011}^{(qijkl)}]
 -[p_{11100}^{(qijkl)}-p_{10011}^{(qijkl)}]
 \\
 \langle Z_q Z_r O_{ijkl}\rangle =&
  [p_{001100}^{(qrijkl)}-p_{000011}^{(qrijkl)}]
 -[p_{011100}^{(qrijkl)}-p_{010011}^{(qrijkl)}]
 -[p_{101100}^{(qrijkl)}-p_{100011}^{(qrijkl)}]
 +[p_{111100}^{(qrijkl)}-p_{110011}^{(qrijkl)}].
 \end{split}
 \end{equation}

In the case where two terms $h_{ijkl}$, $h_{i'j'k'l'}$ are simultaneously diagonalized, the indices from the product of $Z$ matrices in each term might overlap.
If there is an even number of overlapping $Z$ matrices, $P_{ij}^{kl}$ commutes with $M_{i'j'k'l'}$ and the same circuit $M_{i'j'k'l'}$ to diagonalize $h_{i'j'k'l'}$  can be used, since $P_{ij}^{kl}$ can be factored out.
The same holds for $M_{ijkl}$. For example, if there are two overlapping $Z$s,
\begin{equation}
\begin{split}
&M_{ijkl}^\dagger M_{i'j'k'l'}^\dagger \left(Z_{i'}Z_{j'}O_{ijkl}\right)\left(Z_iZ_jO_{i'j'k'l'}\right)M_{ijkl} M_{i'j'k'l'}
\\
&=\left(M_{ijkl}^\dagger O_{ijkl} M_{ijkl}\right)\left(M_{i'j'k'l'}^\dagger O_{i'j'k'l'} M_{i'j'k'l'}\right)
Z_i Z_j Z_{i'} Z_{j'}
\\
&=D_{ijkl} D_{i'j'k'l'}Z_i Z_j Z_{i'} Z_{j'},
\end{split}
\end{equation}
with $D_{ijkl}$ and $D_{i'j'k'l'}$ the corresponding diagonal operators.
If $P_{ij}^{kl}$ contains a product of three $Z$s overlapping with ($i'$, $j'$, $k'$, $l'$),
then two can be factored out so that the problem is reduced to 
simultaneously diagonalizing operators $Z_{l'}O_{ijkl}$ and $Z_{l}O_{i'j'k'l'}$.

\begin{figure}[tb]
    \centering
    \includegraphics[width=0.4\linewidth]{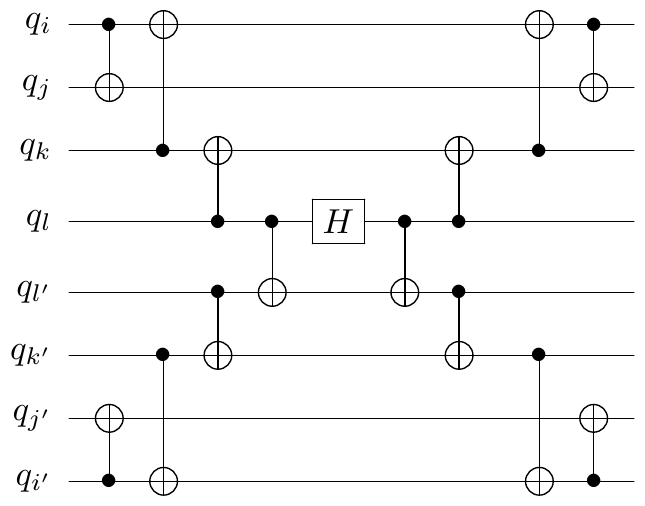}
    \caption{Quantum circuit to implement the change of basis to diagonalize $Z_{l'}O_{ijkl}Z_{l}O_{i'j'k'l'}$
    for double-hopping terms.}
    \label{fig:circuit_overlap}
\end{figure}

In practice, we only need to build new circuits that diagonalize a $2$-qubit subspace, instead of the full $8$-qubit space. The non-diagonal part $O_{ijkl}$ exchanges the states $|0011\rangle$ and $|1100\rangle$, effectively operating in this two-state subspace through an $X$ gate. The circuit in the right dashed box of Fig.~2 in the main text can be interpreted as a three-step protocol. First, a change of basis through a set of CNOT gates such that $X$ operates only in the last qubit; second, a Hadamard gate acting on that qubit to diagonalize $X$, $H X H = Z$; and third, the inverse sequence of CNOTs to switch back  to the original basis. If one term has an overlapping $Z$, then instead of the Hadamard gate acting separately on each $4$-qubit circuit, we need to diagonalize the corresponding $2$-qubit space. For example, if we want to measure $Z_{i'}O_{ijkl}$ and  $Z_{i}O_{i'j'k'l'}$ with the same circuit, we need to diagonalize $X_l Z_{l'}$ and $Z_l X_{l'}$, and embed the corresponding circuit, $CX_{ll'}H_{l'} CX_{ll'}$, within the change of basis, see Fig.~\ref{fig:circuit_overlap}.

\subsection{Circuits to diagonalize products of Hamiltonian and pool operators }\label{app:diagonalization2}

\begin{figure}[tb]
    \centering
    \includegraphics[width=.3\linewidth]{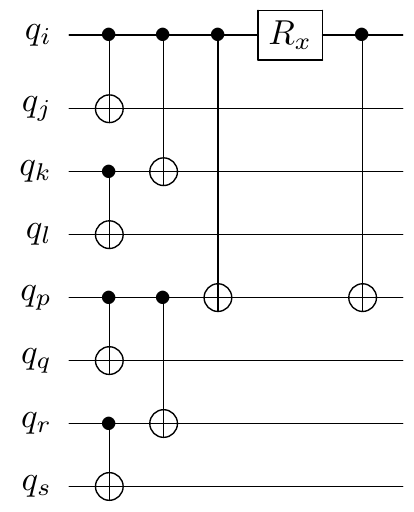}
    \caption{Quantum circuit $M_{ijkl}^{pqrs}$ to diagonalize $h_{ijkl}T_{pq}^{rs}$ when all eight indices are different. The corresponding expectation value, $\langle \psi_n|h_{ijkl}T_{pq}^{rs}|\psi_n\rangle=-p_{00100010}+p_{00101010}-p_{10100010}+p_{10101010}$, depends on $p_m$, the probabilities of measuring $m$ in the corresponding qubits ($i$, $j$, $k$, $l$, $p$, $q$, $r$, $s$) in the statevector $M_{ijkl}^{pqrs}|\psi_n\rangle $.}
    \label{fig:circuit_Mijklpqrs}
\end{figure}

In order to measure gradients using Eq.~(7) in the main text, we need to compute expected values of $h_{ijkl}T_{pq}^{rs}$. Similarly to $O_{ijkl}$, this operator effectively swaps two states in the computational basis,
\begin{equation}
\begin{split}
h_{ijkl}T_{pq}^{rs} &=
i |11001100\rangle\langle00110011| - i |00110011\rangle\langle11001100|
\\&
+ i |00111100\rangle\langle11000011| - i | 11000011\rangle\langle00111100|,
\end{split}
\end{equation}
where we have assumed $P_{ij}^{kl}=P_{pq}^{rs}=1$.
This operator can be disentangled through a series of CNOT gates up to the 2-qubit operator $X_iY_p$, which is then diagonalized with the basis change $CX_{ip}{R_x}_{i}CX_{ip}$. Figure.~\ref{fig:circuit_Mijklpqrs} illustrates the full circuit to diagonalize $h_{ijkl}T_{pq}^{rs}$.

\section*{Discussion on the complete simulation set}\label{app:suppinfo}

We choose the 
Cohen-Kurath interaction~\cite{cohen1965effective} in the $p$ shell, 
USDB~\cite{Brown2006} in the $sd$ shell and 
KB3G in the $pf$ shell~\cite{Poves2001}.
Explicit three-nucleon interactions are typically 
neglected because their leading effects 
can be written as an effective two-body 
term~\cite{Carbone2013,Hebeler:2020ocj}.

\begin{figure*}[tb]
    \centering
    \includegraphics[width=.99\linewidth]{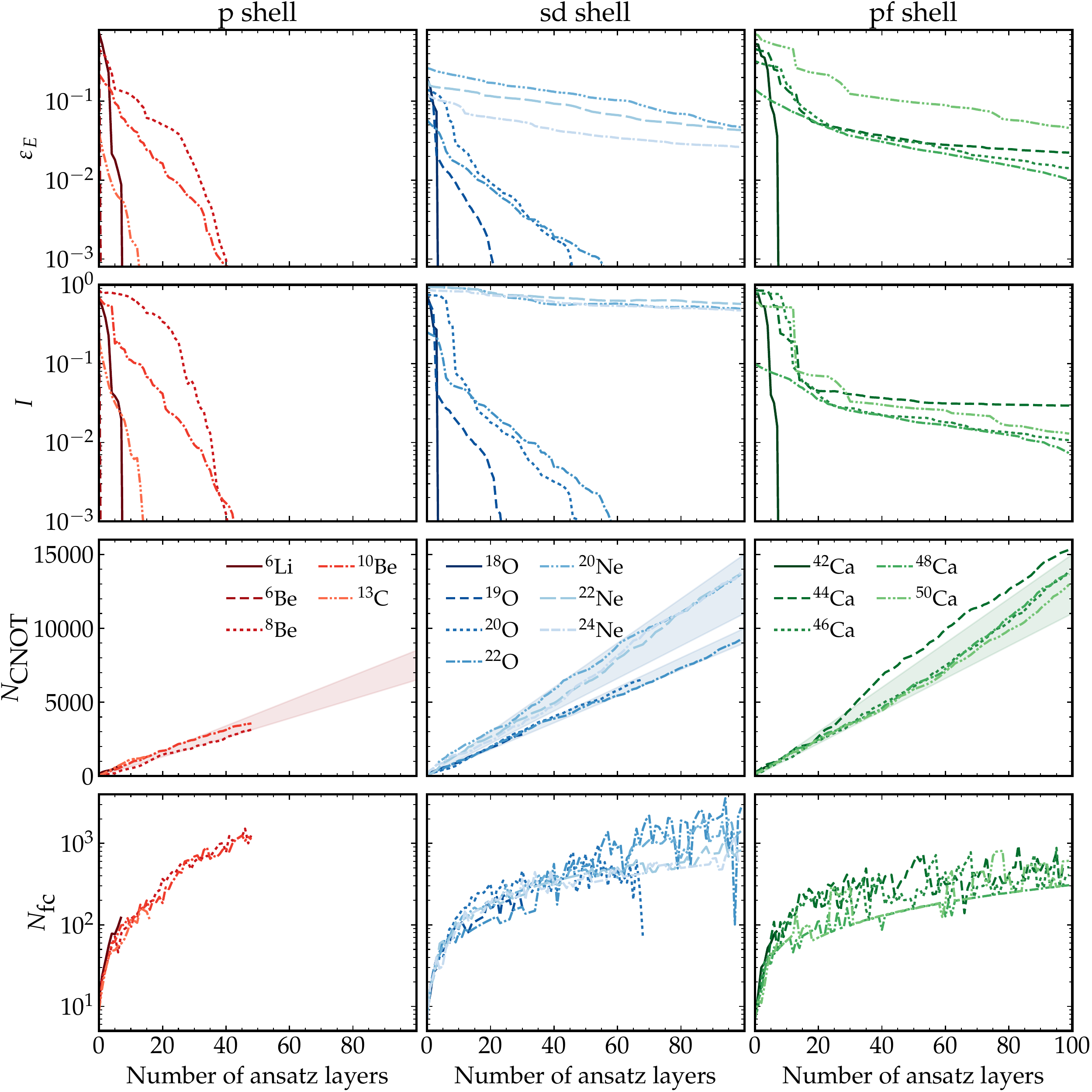}
    \caption{Evolution of the relative error for the ground-state energy, $\varepsilon_{\rm E}$, (top row), infidelity $I$ (second row), number of CNOT gates in the ansatz circuit $N_{\rm CNOT}$ (third row) and number of cost-function calls $N_{\rm fc}$ in the classical optimizer (bottom row) as a function of the number of ansatz layers for simulations of all $p$-shell (first column), $sd$-shell (second column) and $pf$-shell (third column) nuclei considered in this work. The bands in the number of CNOT gates panels are meant to guide the eye and correspond to lower (upper) limits of CNOT gates per layer of 65 (85) in $p$-shell nuclei, 90 (100) in oxygen isotopes and 110 (150) in both neon and calcium isotopes. The number of CNOT gates increases polynomially even in least favorable cases of convergence of $^{44}$Ca and $^{24}$Ne. The relative energy error and infidelities follow analogous trends during the iterative process. This indicates that the algorithm captures the correlations in the nuclear wavefunctions. 
    The number of calls to the cost-function for the classical optimization presents a similar trend for all nuclei, mildly increasing on average with the number of layers.}
    \label{fig:evolution-all-nuclei}
\end{figure*}

Figure~\ref{fig:evolution-all-nuclei} shows the dependence on the number of ansatz layers of the energy error $\varepsilon_{\rm E}$ (top panels), infidelities $I$ (second-row panels), number of CNOTs $N_\textrm{CNOT}$ (third-row panels) and number of cost-function calls used by the classical optimizer $N_\textrm{fc}$ (bottom panels) for all nuclei considered in this work. 
The iterative evolution shown by Fig.~\ref{fig:evolution-all-nuclei} presents similar features to Figs.~4 and ~5 of the main text, where results are shown only for selected nuclei. 

The first column of Fig.~\ref{fig:evolution-all-nuclei} indicates that all nuclei in the $p$ shell are relatively straightforward to implement. They all converge quickly, reaching a relative ground-state energy error $\varepsilon_{\rm E} < 10^{-3}$ with only a dozen layers. Only $^8$Be and $^{10}$Be, with 2 valence protons and 2 and 4 valence neutrons, respectively, require circuit architectures with $\approx$ 50 layers in order to capture their open-shell correlations, converging to a precision below $\varepsilon_{\rm E} = 10^{-6}$. We show $N_\textrm{CNOT}$ and $N_\textrm{fc}$ only up to this point, since this is the accuracy threshold of the classical minimizer. For all cases, the number of CNOT gates increases smoothly with numbers between 65 and 85 gates per layer. Thus, the implementation of $p$-shell nuclei in quantum circuits is promising in terms of both width (number of qubits, $N_{qb}=12$ in this case) and depth (number of total CNOTs). 

Using a single Slater determinant as a reference state is usually enough for the adaptive iterative procedure to reach the ground-state energy and wavefunction exponentially by increasing the number of parameters. 
In some cases, for particularly correlated systems, the initial state may be closer in structure to an excited state than the ground state, and one may land into the local minimum corresponding to the excited state. The only such situation we encountered is $^6$Li, where a simple change of reference state was sufficient to converge into the ground state.
We also note that $^6$Be is represented in the figure, but it converges in only $2$ layers.

The second column of Fig.~\ref{fig:evolution-all-nuclei} shows results for oxygen isotopes  (with no valence protons) and neon (two valence protons) in the $sd$ shell, studied with circuits of $N_{qb}=12$ and $24$ qubits, respectively. We observe a stark difference in the simulation of both isotopic chains: the adaptive procedure ---starting from a single Slater determinant reference state--- needs significantly more layers to capture the many-body correlations present in open-shell neon isotopes. This is due to the relatively large many-body basis dimension of these neon isotopes 
$\dim_{\rm{mb}} \approx 10^{4}-10^{5}$
(see the right panel of Fig. 1 in the main text).
Nevertheless, the number of CNOT gates scales at most polynomially with the number of layers, with between $90$ and $100$ gates per layer for oxygen and between $110$ and $150$ for neon isotopes. This relatively mild non-exponential scaling is promising toward the implementation of ADAPT-VQE in NISQ devices. The bottom panel shows that the number of calls to the cost function used by the classical optimizer at a given iteration is similar for $sd$- and $p$-shell nuclei. This suggests that there is no bottleneck in resources associated to the classical optimizer. 

Finally, the third column of Fig.~\ref{fig:evolution-all-nuclei} presents the results for calcium isotopes (with no valence protons) in the $pf$ shell, using circuits with $N_\textrm{qb}=20$ qubits. 
The first isotope, $^{42}$Ca, convergences extremely 
quickly, within $10$ layers.
In contrast, calcium isotopes with more than $2$ valence neutrons result in a slow convergence, similar to the one for neon isotopes. Again, these calcium isotopes have $\dim_{\rm{mb}} \approx 10^{4}-10^{5}$, and the algorithm needs more updates of the wavefunction to capture the strong correlations in their ground states. We find the slowest convergence for $^{44}$Ca, a midshell isotope between 
the closed-shell $^{40}$Ca and $^{48}$Ca. Likewise, the infidelity of $^{44}$Ca
seems to stall around $I \approx 3 \times 10^{-2}$ 
and even the number of CNOT gates per layer grows beyond the range found for the rest of isotopes. 
This suggests that a different choice of reference state, involving more many-body basis states, may be required for a faster convergence and, as a result, a reduction in quantum resources. In contrast, we find again that the number of cost-function calls for all calcium isotopes follows a similar trend to the $p$- and $sd$-shell nuclei. 

\end{document}